\documentclass[11pt,a4paper]{article}

\pdfoutput=1
\usepackage{jheppub}

\usepackage{braket}

\usepackage{graphicx}
\usepackage{wrapfig,enumerate,slashed}

\usepackage{footmisc}
\usepackage{amsmath}
\usepackage{wasysym} 
\usepackage{color}
\usepackage{comment}

\DeclareMathAlphabet{\mathbold}{U}{zeur}{b}{n}

\usepackage[labelfont=bf,font={small}]{caption}

\renewcommand\[{\left[}
\renewcommand\]{\right]}

\def\beq{\begin{equation}}
\def\eeq{\end{equation}}
\def\[{\begin{equation}}
\def\]{\end{equation}}

\newcommand{\startappendix}{
\setcounter{section}{0}
\renewcommand{\thesection}{\Alph{section}}
\renewcommand{\theequation}{\Alph{section}.\arabic{equation}}}

\newcommand{\Appendix}[1]{
\refstepcounter{section}
\begin{flushleft}
{\Large\bf Appendix: #1}
\end{flushleft}}

\begin{document}
\numberwithin{equation}{section}

\title{Precision measurements for the Higgsploding Standard Model}

\author{Valentin V. Khoze,}
\author{Joey Reiness,}
\author{Michael Spannowsky,}
\author{and Philip Waite}

\affiliation{Institute for Particle Physics Phenomenology, Department of Physics, Durham University, Durham, DH1 3LE, UK}

\emailAdd{valya.khoze@durham.ac.uk}
\emailAdd{joey.y.reiness@durham.ac.uk}
\emailAdd{michael.spannowsky@durham.ac.uk}
\emailAdd{p.a.waite@durham.ac.uk}

\abstract{Higgsplosion is the mechanism that leads to exponentially growing decay rates of highly energetic
particles into states with very high numbers of relatively soft Higgs bosons.
In this paper we study quantum effects in the presence of Higgsplosion.
First, we provide a non-perturbative definition of Higgsplosion as a resolved short-distance singularity of quantum propagators
at distances shorter than the inverse Higgsplosion energy scale, $E_*$. 
We then consider quantum effects arising from loops in perturbation theory with these propagators on internal lines. 
When the loop momenta exceed the Higgsplosion scale $E_*$, the theory dynamics deviates from what is expected in 
the standard QFT settings without Higgsplosion. The UV divergences are automatically regulated by the Higgsplosion scale,
leading to the change of slopes for the running couplings at the RG scales $\mu > E_*$. Thus, the theory becomes 
asymptotically safe. Further, we find that the finite parts are also modified and receive power-suppressed corrections in $1/E_*^2$.
We use these results to compute a set of precision observables for the Higgsploding Standard Model. These and other precision observables could provide
experimental evidence and tests for the existence of Higgsplosion in particle physics.}


\preprint{}

\maketitle



\section{Introduction}
\label{sec:intro}
\medskip

A conventional wisdom is that in the description of nature based on a local quantum field theory, one should always be able to probe 
shorter and shorter distances with higher and higher energies. Specifically, the characteristic length scales probed are
$\Delta x \sim 1/E$ where $E$ is the energy scale, a momentum transfer, or the virtuality that is achieved in the experiment.
In the asymptotic regime where $E\to \infty$, one expects to probe $\Delta x \to 0$.

Higgsplosion \cite{Khoze:2017tjt} is a dynamical mechanism, or a new phase of the theory, which presents an obstacle to this length-scale/energy
principle at energies above a certain value $E_*$, referred to as the Higgsplosion energy scale.
Beyond this energy scale the dynamics of the system is changed drastically \cite{Khoze:2017lft}: distance scales below $|x| \lesssim 1/E_*$ cannot be resolved in interactions; UV divergences are regulated and the theory becomes asymptotically safe.
 This effect can also be depicted in the short-distance scaling behaviour of the propagator of a scalar particle, 

\[
\Delta(x)\,:=\, \langle 0| T(\phi(x)\, \phi(0))|0\rangle\, \sim \, 
\begin{cases}
\,\,m^2\,  e^{-m|x|} & :\,\,{\rm for}\,\,  {|x|} \gg 1/m\\
\,\, 1/|x|^2 & :\,\,{\rm for}\,\,  1/E_* \ll  {|x|} \ll 1/m\\
\,\, E_*^2 & :\,\,{\rm for}\,\,  {|x|}\lesssim 1/E_*
\end{cases}\, ,
\label{prop_Hplosion}
\]
where for ${|x|}\lesssim 1/E_*$ one enters the Higgsplosion regime.

In the simplest settings described by a quantum field theory of a massive scalar field $\phi$ with mass $m$ 
and coupling $\lambda$, we show in Section~\ref{sec:basics} how Eq.~\eqref{prop_Hplosion} is linked to the growing multi-particle decay rates. Furthermore, we show that the Higgsplosion energy scale is set by $E_*\,=\, C\, \frac{m}{\lambda}$, where $C$ is a model-dependent constant of $\mathcal{O}(100)$. This expression holds in the weak-coupling limit $\lambda \to 0$.
In this respect, it resembles the $SU(2)$ sphaleron, which has a mass scale of $M_{\rm sph}\,=\, {\rm const}\, \frac{m_W}{\alpha_w}$ \cite{Manton:1983nd,Klinkhamer:1984di}. However, while the sphaleron is a phenomenon of the non-Abelian gauge-Higgs sector of the Standard Model, Higgsplosion arises due to its scalar sector only.

The fundamental ingredient for the theory is the value of the Higgsplosion scale $E_*$. It is the scale where the rate for the process $1^* \to n \times h$ grows exponentially for large enough $n$. The factorial growth of the rate has been calculated before at leading order \cite{Brown:1992ay,Argyres:1992np,Voloshin:1992rr,Libanov:1994ug}, one-loop resummed \cite{Voloshin:1992nu,Smith:1992rq,Libanov:1994ug,Voloshin:2017flq}, or using a semi-classical approach \cite{Son:1995wz,Libanov:1997nt,Gorsky:1993ix,Khoze:2017ifq}. However, Higgsplosion itself has not been taken into account in those calculations. Thus, in Section~\ref{sec:loops}, we extend their approach by including Higgsplosion, and, for the first time, calculate the loop-corrected rates in a self-consistent way. 

After the Higgsplosion scale $E_*$ is established we can evaluate its phenomenological impact on precision observables, such as $gg \to h^{(*)}$, $h \to \gamma \gamma$, $h \to Z \gamma$, $B \to X_s \gamma$ or $g-2$. We calculate these precision observables explicitly in Section~\ref{sec:qft}, and conclude with a discussion of our findings in Section~\ref{sec:concl}.

\medskip
\section{The propagator and Higgsplosion basics}
\label{sec:basics}

\medskip
\subsection{The Dyson propagator}
\label{sec:Dyson}
\bigskip

In the introduction we pointed out that the central object in a theory with Higgsplosion is the propagator
\eqref{prop_Hplosion},
and that Higgsplosion manifests itself in resolving the short-distance singularity at $x^2 \le 1/E_*^2$,
where $E_*$ is the characteristic (high-)energy scale of Higgsplosion.
To explain what we mean by this and how the effect of Higgsplosion modifies the familiar structure 
of the propagator, it is worthwhile first to summarise the basic elements and the interplay between the propagator 
for a massive scalar field $\phi$, its self-energy $\Sigma(p^2)$, and the partial width $\Gamma_n(p^2)$.
This is the aim of this section.

Our technical discussion in this and the following section will be for a quantum field theory
of a single massive scalar degree of freedom.
The specific models we consider are the $\phi^4$ theory with the unbroken $Z_2$ symmetry,
\[
{\cal L} \,=\,
 \frac{1}{2} \partial^\mu \phi \,\partial_\mu \phi \,-\, \frac{1}{2} m_0^2 \, \phi^2 \,-\,\frac{\lambda}{4} \,\phi^4 \,,
\label{eq:nossb}
\]
and a similar model with spontaneous symmetry breaking (SSB),
where the scalar field has a non-zero VEV $\langle \phi \rangle\,=\,v$,
\begin{eqnarray}
{\cal L} &=&\frac{1}{2}\, \partial^\mu \phi \, \partial_\mu \phi\, -\,  \frac{\lambda}{4} \left( \phi^2 - v^2\right)^2
\label{eq:ssb} \,,\\
&&\phi(x) \,=\, v \,+\, h(x)\, , \quad m_h\,=\, \sqrt{2 \lambda}\, v\,.
\label{eq:vmh}
\end{eqnarray}
In the SSB case \eqref{eq:ssb}-\eqref{eq:vmh}, $m_h$ is the mass of the physical scalar field $h(x)$.
As in our earlier work \cite{Khoze:2017tjt,Khoze:2017ifq,Khoze:2017lft}
we will view the SSB theory \eqref{eq:ssb} as a simplified model for the Standard Model Higgs sector in the unitary gauge. 

\medskip

\noindent In a generic QFT model with a massive scalar, we can define the following quantities:
\begin{enumerate}
\item The Feynman propagator of $\phi$ is the Fourier transform
of the 2-point Green function, 
\[ \Delta (p) \,=\, \int d^4x \, e^{i p\cdot x} \langle 0|T\left( \phi(x) \, \phi(0)\right)|0\rangle \,=\,
\frac{i}{p^2-m_0^2-\Sigma(p^2)+i\epsilon}\,,
\label{eq:dres}
\]
where $m_0$ is the bare (unrenormalised) mass of the scalar field $\phi$.
\item The self-energy $\Sigma(p^2)$ is the the sum of all one-particle-irreducible (1PI) 
diagrams contributing to the 2-point function,
\[
-i\, \Sigma(p^2)\,=\, \sum \,-({\rm 1PI})-\,.
\]
The right-hand side of Eq.~\eqref{eq:dres}
can be interpreted in perturbation theory as the sum over the infinite series of the bare propagators and the $\Sigma(p^2)$ insertions, 
\[
\frac{i}{p^2-m_0^2-\Sigma(p^2)}\,=\, \frac{i}{p^2-m_0^2} \,+\, 
 \frac{i}{p^2-m_0^2}\, \sum_{n=1}^\infty \left(-i \Sigma(p^2)\, \frac{i}{p^2-m_0^2}\right)^n
\,.
\label{eq:drexp}
\]
Hence Eq.~\eqref{eq:dres} gives the full quantum propagator, also known as the Dyson propagator~\cite{Dyson:1949ha, Schwinger:1951ex, Schwinger:1951hq}, valid in perturbative and non-perturbative quantum field theories.
\item The physical (or pole) mass $m$ is defined as the pole of the quantum propagator \eqref{eq:dres},
\[
m^2 - m_0^2 -  \Sigma(m^2)\,=\, 0 \,\,, \quad {\rm or} \quad 
m^2\,=\, m_0^2 \, +\,  \Sigma(m^2)\,.
\label{eq:mdef}
\]
\item The field renormalisation constant $Z_\phi$ is determined from the slope of $\Sigma(p^2)$ at $m^2$,
\[
Z_\phi \,=\, \left(1 -\left.\frac{d \Sigma}{dp^2}\right\vert_{p^2=m^2}\right)^{-1}\,.
\label{eq:mass1}
\]
Using the definition of the pole mass \eqref{eq:mdef} and the renormalisation constant $Z_\phi$ in \eqref{eq:mass1},
the full propagator \eqref{eq:dres} can be written as,
\[\Delta (p)\,=\, 
\frac{i Z_\phi}{p^2-m^2\,-\, Z_\phi [\Sigma(p^2)-\Sigma(m^2)- \Sigma'(m^2) (p^2-m^2)]}\,.
\]
\item The $Z_\phi$ constant is used to define the renormalised quantities $\Delta_R (p)$ and $\Sigma_R(p^2)$,
\begin{eqnarray}
\Delta_R (p) &=& Z_\phi^{\,\,-1} \, \Delta(p)\, , 
\label{eq:RpropS} \\
\Sigma_R (p^2)&=& Z_\phi \,\left(\Sigma(p^2)-\Sigma(m^2)- \Sigma'(m^2) (p^2-m^2)\right)\,.
\label{eq:Sigmasub}
\end{eqnarray}
Hence, the result for the renormalised propagator in terms of all finite quantities is,
\[\Delta_R (p)\,=\, 
\frac{i }{p^2-m^2\,-\, \Sigma_R(p^2) +i\epsilon}\,.
\label{eq:propH}
\]
\item The optical theorem provides the physical interpretation of the imaginary part of the self-energy 
in terms of the momentum-scale dependent decay width $\Gamma(p^2)$,
\[ -\, {\rm Im}\, \Sigma_R (p^2) \,=\,  m\,\Gamma(p^2) 
\label{eq:2.10}
\,,\]
with the decay width being determined by the partial widths of $n$-particle decays at energies
$s \ge (nm)^2$,
\[
\Gamma (s) \,=\, \sum_{n=2}^\infty \Gamma_n (s) \,\,, \qquad
\Gamma_n (s)\,=\,\frac{1}{2m} \int \dfrac{d\Phi_n}{n!}\,  |{\cal M}(1\to n)|^2 \,. \label{eq:2.11}
\]
Here ${\cal M}$ is the amplitude for the $1^*\to n$ process, the integral is over the $n$-particle Lorentz-invariant
phase space, and $1/n!$ is the Bose-Einstein symmetry factor for $n$ spin-zero particles produced in the final state.
\item The origin of Higgsplosion \cite{Khoze:2017tjt} is that the scattering amplitudes ${\cal M}(1\to n)$, and consequentially
the decay rates into the $n$-particle final states, grow factorially with $n$ in the large-$n$ limit, 
$\frac{1}{n!} |{\cal M}_{n}|^2 \sim n! \lambda^n \sim e^{n\log(\lambda n)}$. When $n$ scales linearly with the available energy,
$n \sim \sqrt{s}/m$, this translates into the exponential dependence of the decay rate $\Gamma(s)$ on $\sqrt{s}$.
It was further argued in \cite{Khoze:2017tjt,Khoze:2017ifq} that there is a sharp transition 
between the exponential suppression, $\Gamma_n(s < E_*^2)/m \ll 1$, and the exponential growth, $\Gamma_n(s > E_*^2)/m \gg 1$,
for the $n$-particle rate at a certain characteristic energy scale $E_*$ (and in a large-$n$ limit that is still allowed by kinematics,
 $n \lesssim \sqrt{s}/m$). Hence in a Higgsploding theory, the propagator,
 \[\Delta_R (p)\,=\, 
\frac{i }{p^2-m^2\,-\, {\rm Re}\, \Sigma_R(p^2) \,+\, i m \Gamma (p^2)\, +i \epsilon}\,,
\label{eq:propH2}
\]
is effectively cut off at $p^2 \ge E_*^2$ by the exploding width $\Gamma(p^2)$ of the propagating state into the high-multiplicity
final states. This is the {\it Higgspersion} effect first identified and studied in Refs.~\cite{Khoze:2017tjt,Khoze:2017lft},
which is the direct consequence of Higgsplosion; and
it will play a central role in our discussion of quantum effects in section \ref{sec:loops}. Its physical interpretation is that
the incoming highly virtual state decays rapidly into the multi-particle state made out of soft quanta
with momenta $k_i^2 \sim m^2 \lll E_*^2$. The width of the propagating degree of freedom becomes much greater than its mass
and the virtuality:
it is no longer a simple particle state. In this sense, it has become a composite state made out of the $n$ soft particle quanta 
of the same field $\phi$.
\end{enumerate}

The main purpose of the summary above is to demonstrate that there are no apparent subtleties that arise when
accounting for the UV-renormalisation effects
in the expression for the renormalised propagator \eqref{eq:propH2}. This expression is general and its validity is not
restricted to, for example, the narrow width approximation.

Although we do not have much to say at present about the real part of the self-energy at $p^2 \gg m^2$ (in the regime of
interest where $n\gg1$), it is sufficient for our purposes to consider only the imaginary part, which is determined by the multi-particle 
decay rate $\Gamma_n(p^2)$. As soon as ${\rm Im}\, \Sigma(p^2)\,=-m\, \Gamma_n(p^2)$ becomes exponentially large, 
which occurs
at $E_*$, the  propagators develop sharp exponential form-factors and vanish, thus providing a cut-off above $E_*$
for the integrals over loop momenta. Potential cancellations between the imaginary and real parts are impossible. Essentially, it is sufficient to have the Higgsplosion of the absolute value 
$|\Sigma(p^2)|^2\, =\, ({\rm Re}\, \Sigma(p^2))^2\,+\, ({\rm Im}\, \Sigma(p^2))^2$.

For a more detailed introduction to Higgsplosion and some of its applications we refer the reader to 
Refs.~\cite{Khoze:2017tjt,Khoze:2017ifq,Khoze:2017lft} and \cite{Gainer:2017jkp,Voloshin:2017flq} and references therein.
In particular, we mention here that Higgsplosion solves the fine-tuning problem of the Higgs mass by removing quantum  
contributions to $m_0^2$ from all states with masses or momenta greater than the Higgsplosion scale $E_*$ as explained 
in \cite{Khoze:2017tjt}. We also note the observation of \cite{Khoze:2017lft} that 
no UV divergences remain in the theory above the Higgsplosion scale $E_*$. The running couplings become flat 
above $E_*$ and thus flow to UV fixed points, rendering the theory with Higgsplosion asymptotically safe.

\medskip
\subsection{Continuation to Euclidean space}
\label{sec:Eucl}
\bigskip

In practice, all perturbative calculations in a theory, independently of whether or not it is in the Higgsplosion regime, are 
carried out in Euclidean space. In this paper, we rely on the conjecture that it is valid to analytically continue Higgsplosion into the Euclidean domain. 
In the absence of Higgsplosion, the use of the Euclidean signature $p^2=\,p_0^2 \,+\,\vec{p}^{\,2}$
facilitates the UV regularisation of divergent integrals over loop momenta.\footnote{Dimensional regularisation, cut-off
regularisation and even Pauli-Villars regularisation all are defined and normally carried out in the Euclidean signature.}.
The analytic continuation of the momentum variable is achieved with the standard Wick rotation,
$p^0_{E}= i p_0$ with $\vec{p}_E=\vec{p}$, so that the propagator \eqref{eq:propH} becomes,
\[
\Delta_R (p)\,=\, 
\frac{-i }{p^2+ m^2\,+\, \Sigma_R(p^2) }\,.
\label{eq:propHEu}
\]
In the coordinate representation, using the imaginary time,
\[
\tau = it\,,
\] 
the Dyson propagator reads,
\[
\Delta_R(x_1,x_2) \,=\, \langle 0|\phi(x_1) \phi(x_2)|0\rangle
\,=\, \int \frac{d^4 p}{(2\pi)^4}\,\frac{1}{p^2 +m^2 +\Sigma_R(p^2) }\, e^{i p_0 \Delta \tau \,+\, i\vec{p}\Delta \vec{x}}\,.
\label{eq:propSigma} 
\]
When the theory enters the Higgsplosion regime \cite{Khoze:2017tjt}, the self-energy undergoes a sharp exponential growth. 
This behaviour, 
\[
\Sigma_R(p^2)\,  \sim \, 
\begin{cases}
\,\, 0 \,\, & :\,\,{\rm for}\,\, p^2 < E_*^2 
\\
\,\, \infty & :\,\,{\rm for}\,\,  p^2 \ge E_*^2
\end{cases}\, ,
\label{Sig_Hplosion}
\]
is captured by restricting the integration domain over the loop momenta.
As a result, the integral over the 4-momentum in \eqref{eq:propHEu} 
becomes cut off by $\Sigma$ outside the ball of radius $E_*$,
\[
\Delta_R(x_1,x_2) \,=\,\int_{p^2\le E_*^2} \, \frac{d^4 p}{(2\pi)^4}\,\frac{1}{p^2 +m^2 }\, e^{i p_0 \Delta \tau \,+\, i\vec{p}\Delta \vec{x}}\,.
\label{eq:propEX} 
\]

\noindent For non-vanishing $|\Delta x| > 0$ this integral is regular and can be straightforwardly evaluated.
This leads to short-distance behaviour of the propagator of the form,
\[
\Delta_R(x,0)\, \sim \, 
\begin{cases}
\,\, 1/|\Delta x|^2 & :\,\,{\rm for}\,\,  1/E_* \ll  {|\Delta x|} \ll 1/m\\
\,\, E_*^2 & :\,\,{\rm for}\,\,  {|\Delta x|}\lesssim 1/E_* 
\end{cases}\, .
\label{prop_Hplosion2}
\]
This result is in agreement with Eq.~\eqref{prop_Hplosion}. It is also clear that the absence of the singularity 
at $|x|^2\to 0$ is a consequence of the dynamic integration cut-off
for momenta $p^2 > E_*^2$ (i.e. above the Higgsplosion scale of the self-energy).

\medskip
At the same time, in the opposite limit where the point splitting $\Delta x$ goes to zero, 
we recover from Eq.~\eqref{eq:propEX} 
the loop integral corresponding to the tadpole diagram,
\[
\Delta_R(x,x) \,=\,\int_{p^2\le E_*^2} \, \frac{d^4 p}{(2\pi)^4}\,\frac{1}{p^2 +m^2 }\,.
\label{eq:propLoop} 
\]
It follows that the appearance of the Higgsplosion scale $E_*$ renders this loop integral finite.
In the absence of Higgsplosion, the closed loop integral is quadratically divergent in the UV, as expected, 
and requires UV regularisation either by imposing a UV cut-off $\Lambda_{UV}$, or via dimensional or other type of UV regularisation.
This is an example of the general result that Higgsploding theories are UV-finite \cite{Khoze:2017tjt,Khoze:2017lft}. 

It should be noted that even in the limit of infinite $E_*$ or $\Lambda_{UV}$, the integral in Eq.~\eqref{eq:propEX} is finite for a  non-vanishing end point separation. It is 
is regulated instead by the inverse separation $1/(\Delta t)^2$ or $1/(\Delta \vec{x})^2$. Of course, to obtain a closed loop, we will ultimately have to send this separation to zero.

\medskip
\subsection{Multi-particle decay width}
\label{sec:Gamma}
\bigskip

In the preceding section we explained that the theory enters the Higgsplosion phase if 
the decay rate of a highly virtual / highly energetic single particle state into multi-particle final states 
becomes exponentially large. 
To compute this decay rate $\Gamma_n(s)$, we need to find the amplitudes for the $1^* \to n$ processes 
at energies $\sqrt{s}$ and integrate them over the $n$-particle phase space, as in Eq. \eqref{eq:2.11}.

We are interested in keeping the number of particles $n$ 
in the final state as large as possible, that is, near the maximum number allowed by the phase space, 
$n \lesssim n_{\rm max} = \sqrt{s}/M_h$. We can therefore take the 
bosons in the final state to be non-relativistic.
Such $n$-point amplitudes were studied in detail in scalar QFT 
in Refs.~\cite{Brown:1992ay,Argyres:1992np,Voloshin:1992rr,Libanov:1994ug,Khoze:2014kka}, with the result,
\begin{eqnarray}
{\rm Model}\, \eqref{eq:nossb}:\quad {\cal A}_{1^*\to n} (p_1 \ldots p_n)  &=& 
n!\,  \left(\frac{\lambda}{8m^2}\right)^{\frac{n-1}{2}}\exp\left[-\frac{5}{6}\,n\, \varepsilon \right]\, ,
\label{eq:expsc1}\\
{\rm Model}\, \eqref{eq:ssb}:\quad {\cal A}_{1^*\to n} (p_1 \ldots p_n)  &=& n!\,  \left(\frac{1}{2v}\right)^{n-1}
\exp\left[-\frac{7}{6}\,n\, \varepsilon \right]\label{eq:expsc2}\, , \\
&& n\to \infty\,, \quad \varepsilon \to 0\,, \quad n\varepsilon = {\rm fixed}\,. \nonumber
\end{eqnarray}
As indicated, these tree-level amplitudes are computed in the double-scaling limit with large multiplicities $n \gg 1$ 
and small non-relativistic energies of each individual particle, $\varepsilon \ll 1$, where, 
\[
\varepsilon \,= \, \frac{\sqrt{s}-nm}{nm}\,\,\,=\,\,\,\, \frac{1}{n m}\, E_n^{\rm \, kin}\, \simeq\, 
\frac{1}{n}\,  \frac{1}{2 m^2} \, \sum_{i=1}^n \vec{p}_i^{\, \, 2} \,,
\label{eq:epsdef}
\]
so that the total kinetic energy per particle mass $n\varepsilon$ in the final state is fixed.

The pre-exponential factors on the right-hand side of Eqs.~\eqref{eq:expsc1}-\eqref{eq:expsc2}  correspond to the tree-level amplitudes 
(or more precisely, currents with one incoming off-shell leg) computed on
the $n$-particle thresholds. In the elegant formalism of Brown \cite{Brown:1992ay}, 
these amplitudes for all $n$ arise from the generating functional $\phi_0$, which is given
by a spatially-uniform solution to the classical equations of 
motion.\footnote{Tree level is equivalent to the leading-order expansion in $\hbar$, thus reducing the
quantum problem to classical solutions.
Furthermore, the vanishing external 3-momenta
of the on-the-threshold amplitudes selects spatially-uniform time-dependent solutions \cite{Brown:1992ay}.} 
These solutions are easily found by solving the classical equations of
motion corresponding to models \eqref{eq:nossb}-\eqref{eq:ssb}, and read,
\begin{eqnarray}
{\rm Model}\, \eqref{eq:nossb}:\qquad 
\phi_0(t)   &=& 
\frac{z(t)}{1-\lambda z(t)^2/(8m^2)}\,, \quad {\rm where} \quad
z=z_0 \, e^{i m t}
\label{eq:sol_nossb}\\
{\rm Model}\, \eqref{eq:ssb}:\qquad 
\phi_0(t)   &=& 
v\left( \frac{1+z(t)/(2v)}{1+z(t)/(2v)}\right)\,, \quad {\rm where} \quad
z=z_0 \, e^{i m_h t}
\label{eq:sol_ssb}
\end{eqnarray}
Then the amplitudes on $n$-particle mass thresholds are given by differentiating these classical generating
functions $n$ times with respect to the source variable $z$, 
\[
{\cal A}^{\rm thr.}_{\rm tree\,\, 1^*\to n}\,:=\,  \langle n|\phi |0\rangle_{\rm tree}\,=\, 
\left(\frac{\partial}{\partial z}\right)^{n} \phi_0 (z) \, |_{z=0}\,=\,
\begin{cases}
\,\, n!\,  \left(\frac{\lambda}{8m^2}\right)^{\frac{n-1}{2}}
\, & 
\\
\,\, n!\,  \left(\frac{1}{2v}\right)^{n-1} 
\,=\,n!\,\left(\frac{\lambda}{m_h^2}\right)^\frac{n-1}{2}\,.
\, & 
\end{cases}\hspace{-1cm}
\label{eq:expsc12}
\]
These are
exact expressions for tree-level amplitudes valid for any value of $n$ \cite{Brown:1992ay}.

The amplitudes in Eqs.~\eqref{eq:expsc1}-\eqref{eq:expsc2} go beyond the threshold results \eqref{eq:expsc12} by
accounting for the  
kinematic dependence 
in the non-relativistic limit of the external momenta. These were derived as a non-relativistic deformation 
of the Brown's generating functions method in Refs.~\cite{Libanov:1994ug} and \cite{Khoze:2014kka} for the unbroken and
the broken theories respectively.
The amplitudes now contain an exponential form-factor that depends on the kinetic energy of the final state $n\varepsilon$.
But, importantly, the factorial growth $\sim \lambda^{n/2}\,  n!$ characteristic to the multi-particle amplitude on mass threshold remains.
Its occurrence can be traced back to the factorially growing number of Feynman diagrams at large $n$
and the lack of destructive interference 
between the diagrams in the scalar theory.
In Section~\ref{sec:loops} we will compute the leading-order quantum corrections to these tree-level amplitudes, 
in the case where a non-trivial finite Higgsplosion scale $E_*$ is present.

To obtain the self-energy, ${\rm Im}\, \Sigma_R (p^2) \,=\,  -m\,\Gamma(p^2)$, we
integrate the amplitudes\footnote{Note that the conventionally-normalised 
amplitudes ${\cal M}_{1^*\to n}$ are obtained from the currents ${\cal A}_{1^*\to n}$ 
in \eqref{eq:expsc1}-\eqref{eq:expsc2}  by the LSZ amputation 
of the single off-shell incoming line,
${\cal M}_{1\to n}  \,: =\, (s-M_h^2) \,\cdot\,{\cal A}_{1^*\to n} (p_1 \ldots p_n)$.}
\eqref{eq:expsc1}-\eqref{eq:expsc2} over the $n$-particle phase-space for large $n$,
\[
\Gamma_n (s)\,=\,\frac{1}{2m} \int \dfrac{d\Phi_n}{n!}\,  |{\cal M}_{1\to n}|^2 \,. \label{eq:Rdef}
\]
For the unbroken theory \eqref{eq:nossb}, one finds \cite{Son:1995wz,Libanov:1997nt},
\[
\label{eq:Rtreesimp1}
\Gamma_n(s) \, \propto\,   {\cal R}(\lambda; n,\varepsilon)\,=\, 
\exp \left[\, n\, \left(\log \frac{\lambda n}{16} \,+\, \frac{3}{2}\log \frac{\varepsilon}{3\pi} \,+\, \frac{1}{2} 
\, -\,\frac{17}{12}\,\varepsilon \,+\, Q(\lambda n,\varepsilon)\right) \right]\,,
\]
and for the model with SSB \eqref{eq:ssb} \cite{Khoze:2014kka}, 
\[
\label{eq:Rtreesimp2}
\Gamma_n(s) \, \propto\,   {\cal R}(\lambda; n,\varepsilon)\,=\, 
\exp \left[\, n\, \left(\log \frac{\lambda n}{4} \,+\, \frac{3}{2}\log \frac{\varepsilon}{3\pi} \,+\, \frac{1}{2} 
\, -\,\frac{25}{12}\,\varepsilon \,+\,Q(\lambda n,\varepsilon)\right) \right]\,.
\]
In particular, note that the ubiquitous factorial growth of the large-$n$ amplitudes  translates into the 
 $\frac{1}{n!} |{\cal M}_{n}|^2 \sim n! \lambda^n \sim e^{n\log(\lambda n)}$ factor in the rates ${\cal R}$ above.
The term $Q(\lambda n,\varepsilon)$ on the right-hand side of both equations above indicates
quantum corrections: at tree level $Q(\lambda n,\varepsilon)\,\equiv 0$.

These rates can also be computed using an alternative semi-classical method formulated by Son in Ref.~\cite{Son:1995wz}.
This is an intrinsically non-perturbative approach, with no reference in its outset made to perturbation theory. The path integrals contributing to the amplitudes and the rates are computed in the steepest descent method,
controlled by two large parameters, $1/\lambda \to \infty$ and $n\to \infty$. More precisely, the limit is,
\[ \lambda \to 0\,, \quad n\to \infty\,, \quad {\rm with}\quad
\lambda n = {\rm fixed}\,, \quad \varepsilon ={\rm fixed} \,.
\label{eq:limit}
\]

The semi-classical computation carried out in \cite{Son:1995wz}, in the regime where,
\[ 
\lambda n = {\rm fixed} \ll 1 \,, \quad \varepsilon ={\rm fixed} \ll 1 \,,
\label{eq:limit3}
\]
provided an alternative derivation of the tree-level perturbative results for non-relativistic final states
\eqref{eq:Rtreesimp1}-\eqref{eq:Rtreesimp2}, with the expressions being correctly reproduced.
Remarkably, this semi-classical calculation also reproduces 
the leading-order quantum corrections arising from resumming one-loop effects. These were previously computed
perturbatively in the $\varepsilon \to 0$ limit
in \cite{Voloshin:1992nu,Smith:1992rq,Libanov:1994ug},
\[
Q_{\rm 1-loop}(\lambda n) \,=\,
\begin{cases}
\,\, -\, \lambda n\,\,{\rm Re}\,F/8
\, & :\,\, {\rm Model}\, \eqref{eq:nossb}
\\
\,\, +\, \lambda n\,\,2B
\, & :\,\, {\rm Model}\, \eqref{eq:ssb}
\end{cases},
\label{eq:FB}
\]
where the constant coefficients $F$ and $B$ will be given in Eqs.~\eqref{eq:Fnossb} and \eqref{eq:finB}
below. This agreement constitutes an important test of the applicability of the semi-classical method of Son for
computations of quantum corrections. In Section~\ref{sec:loops} we will verify that this agreement persists
when the perturbative loop integrals are carried out in the presence of the Higgsplosion scale $E_*$.

The semi-classical approach is equally applicable and more relevant to the realisation of  
the non-perturbative Higgsplosion case where,
\[ 
\lambda n = {\rm fixed} \gg1 \,, \quad \varepsilon ={\rm fixed} \ll 1 \,.
\label{eq:limit2}
\]
This calculation was carried out in Ref.~\cite{Khoze:2017ifq} for the spontaneously broken theory
\eqref{eq:ssb}, with the result given by,
\[
{\cal R}_n(\lambda;n,\varepsilon)\,= \, 
\exp \left[ \frac{\lambda n}{\lambda}\, \left( 
\log \frac{\lambda n}{4} \,+\, 0.85\, \sqrt{\lambda n}\,+\,\frac{1}{2}\,+\,\frac{3}{2}\log \frac{\varepsilon}{3\pi} 
\, -\,\frac{25}{12}\,\varepsilon 
\right)\right] ,
\label{eq:Rnp2}
\]
which is equivalent to Eq.~\eqref{eq:Rtreesimp2} with the substitution,
\[
Q_{\lambda n \gg 1}(\lambda n) \,=\, +\, 0.85\, \sqrt{\lambda n}\,.
\label{eq:FB2}
\]
This result \eqref{eq:FB2}~\cite{Khoze:2017ifq} also agrees with an earlier calculation of the semi-classical
amplitude on threshold in Ref.~\cite{Gorsky:1993ix}.

Higgsplosion becomes operative when the decay width, or equivalently the rate \eqref{eq:Rnp2},
becomes exponentially large. Let us estimate the energy $E_*$ where this occurs.
The expression \eqref{eq:Rnp2} was derived in the near-threshold limit \eqref{eq:limit2}, where the parameter $\varepsilon$ 
is treated as a fixed constant much smaller than one. The initial state energy and the final state multiplicity are related linearly via
$ E/m_h \,=\, (1 + \varepsilon) \, n$. Hence, for any fixed $\varepsilon$, one can raise the energy to 
obtain an arbitrarily large $n$, and consequently, a large $\sqrt{\lambda n}$.
The negative $\log \varepsilon$ term and the positive $\sqrt{\lambda n}$ term 
in \eqref{eq:Rnp2} are in competition. Nonetheless, there is always a range of sufficiently high multiplicities, where $\sqrt{\lambda n}$
overtakes the logarithmic term $\log \varepsilon$ for any fixed (however small) value of $\varepsilon$.
This leads to the exponentially growing multi-particle rates. It was shown in Ref.~\cite{Khoze:2017tjt} that in the model
where the multi-particle rates are given by the expression \eqref{eq:Rnp2}, and fixing $\lambda = 1/8$,
the rates start growing exponentially
when $E_*\sim 200 m_h$.

\medskip
\subsection{Where is Higgspersion in the semi-classical calculation?}
\label{sec:higgspersion}
\bigskip

To compute the $n$-particle decay width $\Gamma_n (E)$ in Eq.~\eqref{eq:Rdef} one starts with the amplitude
${\cal M}_{1\to n} $ between an off-shell single particle state and an $n$-particle on-shell final state.
The amplitude itself is obtained from the $n$-particle form factors of generic local operators $\mathcal{O}(x)$
in the theory. The form factors are defined as matrix elements of the operator $\mathcal{O}(x)$ at the origin, $x=0$,
between the vacuum in-state and the $n$-particle out-state
$|n\rangle^{\rm out}_E$ of the fixed total energy $E$ (or more precisely $p_{\rm fin}^2=E^2$),
\[
F_{\mathcal{O}} (n;E)\,=\, {}^{\rm in}\langle 0|\mathcal{O}(0)|n\rangle^{\rm out}_E\,.
\label{eq:formfactor}
\]
The local operator $\mathcal{O}(0)$ creates an off-shell state at the spacetime point $x=0$
from the in-state vacuum. The virtuality of this state is $p^2=E^2$
which is guaranteed by the energy conservation in the transition process to the final state.

The $1^*\to n$ amplitude of interest is obtained from the form factor \eqref{eq:formfactor}
for the operator $\mathcal{O}(x)=\phi(x)$ by LSZ amputating the single incoming $\phi$-leg,
\[
(\mathrm{LSZ})^1_{\phi} \cdot F_\phi (n;E)\,, \quad {\rm where} \quad   
F_\phi (n;E)\,=\, {}^{\rm in}\langle 0|\phi(0)|n\rangle^{\rm out}_E\,.
\label{eq:ffphi}
\]
On the other hand, as was explained in \cite{Son:1995wz}, the semi-classical formalism for computing the $n$-particle rates
cannot be used when the operator is the elementary field, $\mathcal{O}(x)=\phi(x)$. Instead the formalism was developed  in \cite{Son:1995wz}
for the case when the initial off-shell state was created by a composite local operator.
The operator used in \cite{Son:1995wz} was an exponential, $\mathcal{O}(x)=e^{j\phi(x)}$, so that one can obtain the semi-classical approximation
for the $n$-particle rates of the type,
\[
{\cal R}_n(E) \,=\, \lim_{j\to 0} \int \frac{d\Phi_n}{n!}\, \left| {}^{\rm in}\langle 0|e^{j\phi(0)}|n\rangle^{\rm out}_E \right|^2\,=\,
 \lim_{j\to 0} \int \frac{d\Phi_n}{n!}\, \left| F_{\mathcal{O}} (n;E) \right|^2
\,.
\label{eq:RneS}
\]
The semi-classical approximation to \eqref{eq:RneS} is the expression in Eq.~\eqref{eq:Rnp2}
derived in \cite{Khoze:2017ifq} and valid in the large-$n$ regime \eqref{eq:limit2}.

There are two questions we would like to comment on: 1) How does the semi-classical computation of the `wrong' observable
$\propto \left| F_{\mathcal{O}} (n;E) \right|^2$ relate to the original observable $\propto \left| F_\phi (n;E) \right|^2$, which corresponds to the transitioning of the 
`physical' $1$-particle state to $n$ final-state particles? 2) How does Higgspersion impact on the semi-classical rate \eqref{eq:RneS} and \eqref{eq:Rnp2}?

To answer the first question we quote  \cite{Son:1995wz} in stating that the current semi-classical approximation 
can be trusted only to an exponential accuracy, and the dependence of the final result on the specific form of the
operators $\mathcal{O}(x)$ used in the calculation affects only the factor in front of the exponent in \eqref{eq:Rnp2}.
Essentially, this statement is a conjecture of the underlying approach \cite{Rubakov:1991fb,Son:1995wz} --- it was partially 
evidenced in Ref.~\cite{Libanov:1995gh}  but still remains a conjecture. 
We do not have a new insight on this 
quite well-known technical problem, and will assume that the semi-classical computation of the rate in \eqref{eq:RneS}
does give correct information about the exponential form of the physical rate.

There is however an important point relevant for this conjecture of Son and others that we want to emphasise. 
As we shall explain now, the semi-classical equivalence
between observables made out of the operators ${\cal O}(x)$ and the operators $\phi (x)$ is between the {\it decay rates} of the appropriate initial 
states into $n$-particle states, not between the form factors of  ${\cal O}(x)$ and $\phi (x)$ themselves. 
The definition of the decay rate involves matrix elements where all external propagators have been LSZ-amputated, so in particular,
the initial state is non-propagating. Hence the semi-classical equivalence is between,
\[
(\mathrm{LSZ})^1_{\phi} \cdot  {}^{\rm in}\langle 0|\phi(0)|n\rangle^{\rm out}_E
\qquad {\rm and} \qquad {}^{\rm in}\langle 0|\mathcal{O}(0)|n\rangle^{\rm out}_E\,,
\label{eq:equivLSZ}
\]
where the factor $(\mathrm{LSZ})^1_{\phi}$ implies the LSZ amputation of the external propagator of the in-state
$ {}^{\rm in}\langle 0|\phi(0)$. At the same time, there is no such factor on the right-hand side since, as we will see momentarily, 
the state ${}^{\rm in}\langle 0|\mathcal{O}(0)$ is already LSZ-reduced.

One way to see this is to imagine extending the theory
by adding a source term to its Lagrangian and a kinetic term for the new degree of freedom,
\[
\int d^4 x\, {\cal L}[\phi(x)]  \,+\, \int d^4 x \, s(x)\,{\cal O}(x)\,-\, \int d^4x \, s(x) \,{\cal K} \,s(x)\,.
\]
Here $s(x)$ is a field describing a new degree of freedom, 
${\cal O}(x)$ --- a composite operator made out of $\phi$ --- 
is its source coupled to the field $s(x)$, and ${\cal K}$ is the kinetic operator acting on $s$, i.e. the inverse $s$-propagator,
$ \partial_\mu \partial^\mu + M_s^2$.

 \begin{figure}[t]
\begin{center}
\begin{tabular}{c}
\includegraphics[width=0.4\textwidth]{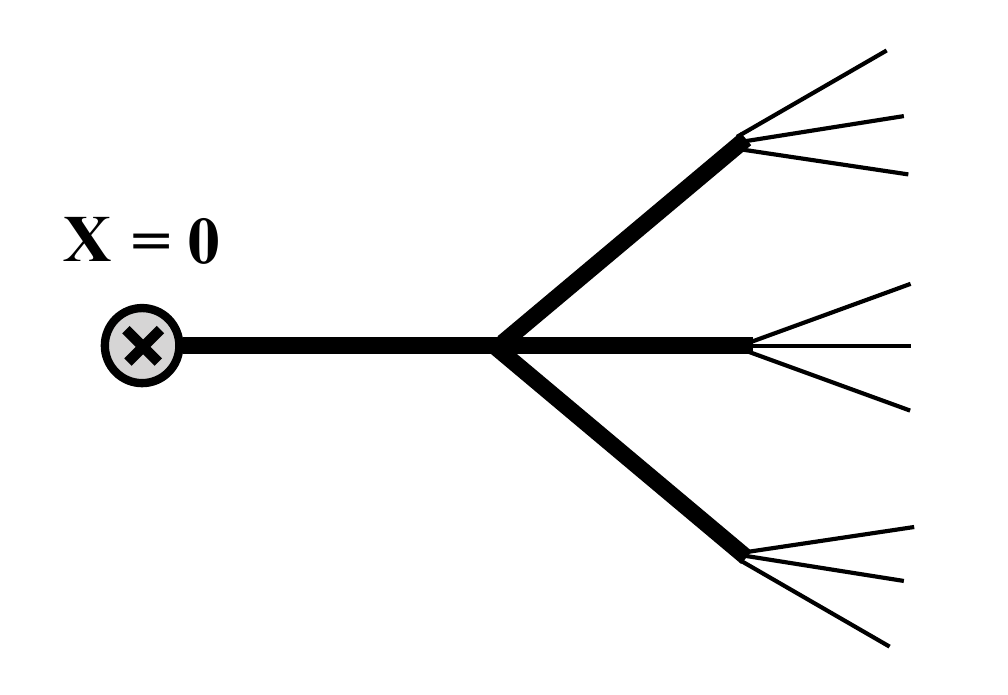} \qquad \quad
\includegraphics[width=0.3\textwidth]{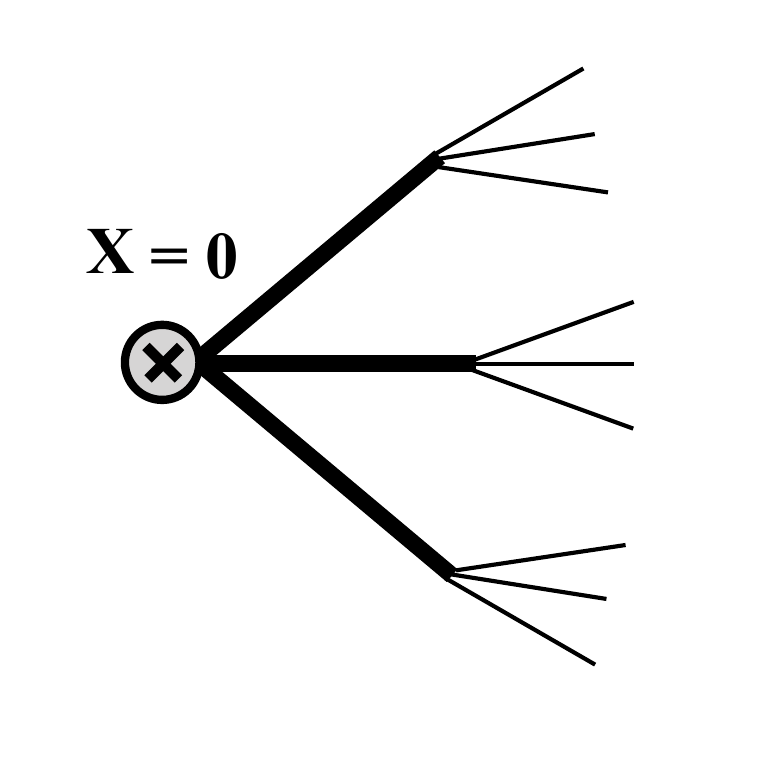}
\\
\end{tabular}
\end{center}
\vskip-.6cm
\caption{
The form factor $F_\phi (n;E)$ of the elementary field $\phi(0)$ is shown in the left diagram, versus
the form factor $F_{\mathcal{O}} (n;E)$ of the composite local operator $\mathcal{O}(0)=\phi^3(0)$, depicted on the right.
Thick lines denote full propagators, thin lines indicate the on-shell $n$-particle final state (in this case $n=9$) with the corresponding propagators already removed.
}
\label{fig:phi}
\end{figure}

The matrix elements to be compared are,
\[ 
{}^{\rm in}\langle 0|\phi(0)|n\rangle^{\rm out}_E 
\qquad {\rm and} \qquad
{}^{\rm in}\langle 0|s(0)|n\rangle^{\rm out}_E\,,
\]
both containing the incoming propagator for the initial state created by either $\phi$ or $s$ from the vacuum.
After LSZ-reducing the incoming propagator in each case, we obtain 
\[ 
(\mathrm{LSZ})^1_{\phi} \cdot {}^{\rm in}\langle 0|\phi(0)|n\rangle^{\rm out}_E 
\qquad {\rm and} \qquad
(\mathrm{LSZ})^1_{s} \cdot {}^{\rm in}\langle 0|s(0)|n\rangle^{\rm out}_E\,=\, {}^{\rm in}\langle 0|{\cal O}(0)|n\rangle^{\rm out}_E\,.
\]
${\cal O}(x)$ is the source term of the field $s(x)$, i.e. 
${\cal K}\cdot s(x)\,=\, {\cal O}(x)$, and it no longer contains the incoming propagator which has been LSZ-amputated.
(We note that the role of the composite operator ${\cal O}(x)$ as the source ${\cal J}$ for a new degree of freedom $s(x)$ and 
the appearance of ${\cal O}(x)$ in the LSZ-amputated amplitudes is completely analogous to the Witten diagrams \cite{Witten:1998qj} in 
the context of the AdS/CFT correspondence.)
Eq.~\eqref{eq:equivLSZ} follows from this observation. The decay rates computed from these two matrix elements are supposed to be 
equivalent within the semi-classical approximation of \cite{Son:1995wz} but we note that we are comparing the
pure form factor of the composite operator ${}^{\rm in}\langle 0|{\cal O}(0)|n\rangle^{\rm out}_E$ to the 
form factor $(\mathrm{LSZ})^1_{\phi} \cdot {}^{\rm in}\langle 0|\phi(0)|n\rangle^{\rm out}_E $ acted on by the LSZ operator.
Without the LSZ reduction, the form factor of the elementary field ${}^{\rm in}\langle 0|\phi(0)|n\rangle^{\rm out}_E $ would be
suppressed by the effect of the self-energy in the external propagator --- the Higgspersion effect \cite{Khoze:2017tjt}.

We can also see this difference between the form factors for the elementary field operator $\phi(x)$ and for the 
local composite operators $\mathcal{O}(x)$ diagrammatically.
A Feynman diagram representation of the form factor $F_\phi (n;E)$ in \eqref{eq:ffphi} is shown on the left diagram in Fig.~\ref{fig:phi}.
We note that this form factor  contains the propagator connecting the point $x=0$ where the fundamental field operator
$\phi(0)$ is located, to the first interaction vertex. The propagator is the full propagator \eqref{eq:propH2}
of the interacting theory,
 \[
 \Delta (p)\,=\, 
\frac{i }{p^2-m^2\,-\, \Sigma(p^2) +i \epsilon}\,,
\label{eq:propH21}
\]
and since it includes the self-energy contribution,  it automatically accounts for the effect of Higgspersion at the momentum scale
where $\Sigma(p^2) \gtrsim p^2$.
On the other hand, diagrams for {\it composite local} operators $\mathcal{O}(x)$ do not have an incoming propagator.
It is easily seen
from the right diagram in Fig.~\ref{fig:phi} which corresponds to $\mathcal{O}(0)=\phi^3(0)$ (but can equally well be used for
any composite operator $\mathcal{O}(0)=\phi^n(0)$) that the point $x=0$ is the vertex, rather than the end point of the in-state propagator associated with 
$\phi(0)$.
Thus, the form factors of composite operators used in the semi-classical calculation do not include the Higgspersion effect 
as there is no in-state propagator, while the form factor of the elementary field does account for Higgspersion. 

\begin{figure}[t]
\begin{center}
\includegraphics[width=0.8\textwidth]{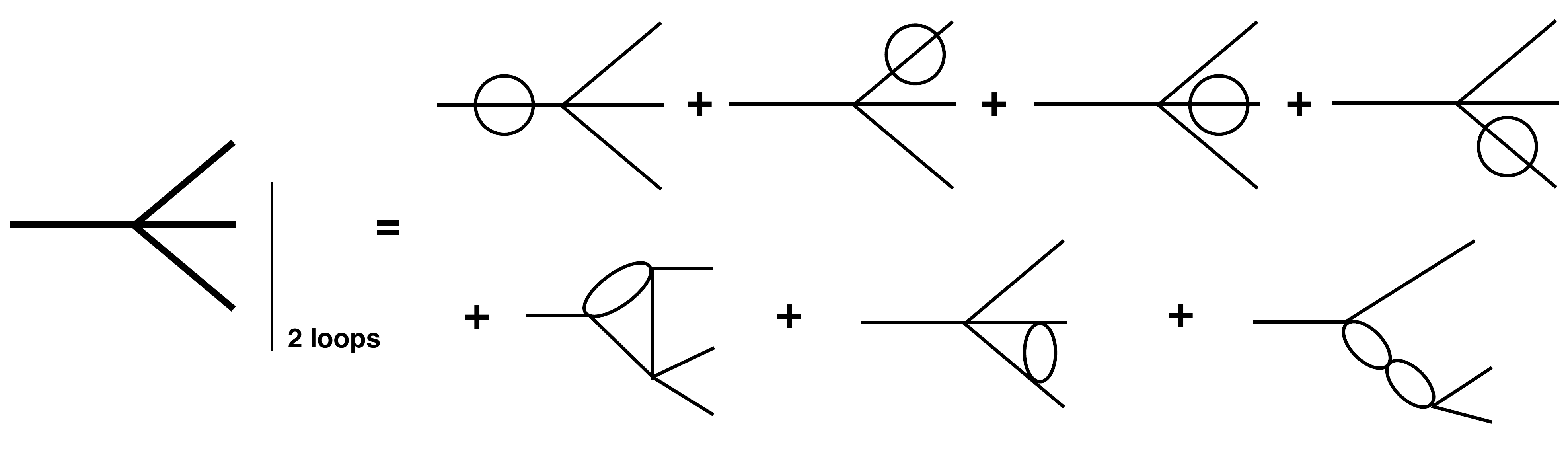}
\end{center}
\caption{The form factor  $F_\phi (n;E)$ of the elementary field $\phi(0)$ at 2-loop order.}
\label{fig:2loop}
\end{figure}
\begin{figure}[t]
\begin{center}
\includegraphics[width=0.8\textwidth]{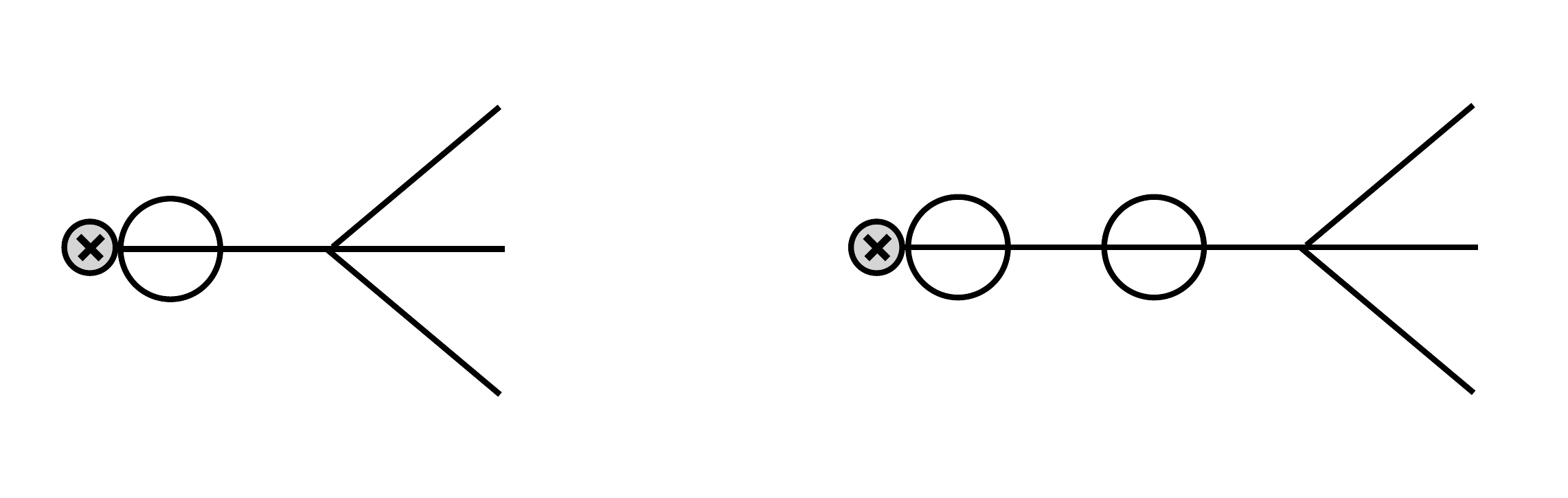}
\end{center}
\caption{Contributions to the form factor  $F_{\phi^3} (n;E)$ suppressed by Higgspersion due to the appearance of the single intermediate propagator. For more details see also \cite{Khoze:2017lft}.}
\label{fig:2loop3}
\end{figure}

One can perhaps better visualise the difference between the form factors in Fig.~\ref{fig:phi} by considering  the 2-loop 
perturbative contribution to the form factor of the elementary field $\phi$. It is shown in Fig.~\ref{fig:2loop} for the $\phi^4$ theory.
Clearly, there are self-energy corrections to the incoming propagator (shown in the first diagram on the right-hand side)
thus resulting in this propagator becoming of the form  \eqref{eq:propH21} after the resummation. On the other hand, there are
no such contributions for the form factors of composite operators, such as shown in the diagram on the right of Fig.~\ref{fig:phi}.
Of course one can draw some one-particle-reducible diagrams contributing to the form factors of composite operators,
such as the ones shown in Fig.~\ref{fig:2loop3}, but it is easy to see that at energies $\sqrt{s} \sim E_*$ their contributions are negligible 
relative to those on the right diagram in Fig.~\ref{fig:phi}. The reason is that the diagrams in Fig.~\ref{fig:2loop3} are exponentially suppressed 
by Higgspersion in the single intermediate propagator at $p^2 = s \gtrsim E_*^2$, while the diagrams on the right plot in Fig.~\ref{fig:phi}
are not. In the latter case the dominant contributions come from the energies and momenta split roughly in 3, hence the virtuality scale
in each of the three branches would be $\sim E_*/3$ which makes the Higgspersion effect completely negligible. 
Similar considerations were explored in more detail in an earlier publication \cite{Khoze:2017lft}.

\bigskip

Of course, as we have already explained,
for the calculation of the actual decay rates $\Gamma_n$,  in each case one needs to 
start from the matrix elements with the LSZ-reduced external lines. 
In other words, the incoming line for the form factor on the left figure
in Fig.~\ref{fig:phi} should be LSZ amputated in accordance with \eqref{eq:ffphi}, while the form factor of the composite operator 
on the right diagram in Fig.~\ref{fig:phi} is already without the external propagator.
Thus, the actual decay rate $\Gamma_n (E) \propto {\cal R}_n(E)$ does not contain the Higgspersing propagator for the in-state
in either case. As the semi-classical calculation in  \cite{Son:1995wz,Khoze:2017ifq} was based on form factors of composite operators,
it did not require an additional LSZ amputation of the incoming full propagator as it was not there in the first place.
The rate in \eqref{eq:RneS} is the imaginary part of the self-energy $\Sigma(p^2)$ in \eqref{eq:propH21}; it is not the 
imaginary part of the whole propagator \eqref{eq:propH21}.

If it was possible to directly compute the form factors of the fundamental operators $F_\phi (n;E)$ in some alternative formalism,
rather than working with the composite operator form factors $F_O (n;E)$,
one would then have to LSZ amputate the external propagators from the integral,
\[
(\mathrm{LSZ})^2_\phi \cdot   \int \frac{d\Phi_n}{n!}\, \left| F_\phi (n;E) \right|^2
\,,
\label{eq:RneSphi}
\]
as an additional step, but as we have explained above, this is not needed for our composite operator-based WKB derivation.

It is only when we want to consider a physical scattering process of $2$ to $n$ particles that we need to restore the propagator \eqref{eq:propH21} of the off-shell state $1^*$. Hence, we conclude that the fully LSZ-amputated decay rates
$\Gamma_n$ can Higgsplode, but the physical scattering processes will always include the Higgspersing propagator and will be 
well behaved at high energies and consistent with unitarity.

\subsection{The Higgsplosion scale $E_*$}
\label{sec:higgspersionE}
\bigskip

In general, the Higgsplosion scale $E_*$ is defined as the new dynamically generated scale of the 
weakly coupled theory, $\lambda \ll 1$, 
where a sharp transition occurs between the negligibly small multi-particle rates at $E<E_*$, and the exponentially 
large rates at $E>E_*$. From general principles, it is clear that this scale would have the parametric dependence of the form,
\[
E_*\,=\, \frac{m_h}{f(\lambda)} \   , \qquad {\rm where} \quad
f(\lambda)|_{\lambda \to 0} \,\to\, 0 \,.
\label{eq:EHgendef}
\]
The mass $m_h$ is the only dimensionful parameter of the theory and the fact that $E_*$ goes to infinity 
in the free-theory limit $\lambda\to 0$ is also obvious since in this limit no multi-boson production is possible.
Hence, the singularity at $\lambda = 0$ implies that the Higgsplosion scale is non-perturbative.

To sharpen this argument for the emergence of $E_*$ (as the new dynamical scale in a weakly coupled theory with
microscopic massive scalars), we would like to get a better handle on $f(\lambda)$ in the context of the 
model with spontaneously broken symmetry \eqref{eq:ssb}. The rate, in this case, is given in Eq.~\eqref{eq:Rnp2}.
It will be useful to introduce the more natural rescaled variables for the multi-particle rate ${\cal R}$,
\[ 
\tilde{n}\,=\, \lambda n \,,
\qquad \tilde{E}\,=\, \frac{\lambda E}{m_h}\,,
\label{eq:tildeEn}
\]
so that 
\[
\tilde{E} \,=\, (1+\varepsilon)\, \tilde{n}\,.
\label{eq:tildelin}
\]

Following closely the discussion in Section 5 of Ref.~\cite{Khoze:2017ifq},
we now re-write the expression for the rate
in the following form,
\[
{\cal R}_n(E)\,\sim \,  \int_0^{\varepsilon_{nr}} d \varepsilon\, \left(\frac{\varepsilon}{3 \pi}\right)^{\frac{3n}{2}} 
\exp \left[ \frac{\tilde{n}}{\lambda}\, \left( 
0.85\, \sqrt{\tilde{n}} \,+\, \log \tilde{n}\,-\, a\, \varepsilon \, \tilde{n}^{\,p}
\,-\, {\rm const}
\right)\right] \,.
\label{eq:RnpN}
\]
The integral $\int^{\varepsilon_{nr}} d \varepsilon\, \left(\frac{\varepsilon}{3 \pi}\right)^{\frac{3n}{2}} $ appearing on the right-hand side
is nothing but the non-relativistic phase-space integral $\int d\Phi_n$ written in the large-$n$ limit. The upper limit
of the integration, $\varepsilon_{nr}$, is the non-relativistic kinetic energy per particle per mass, 
$\varepsilon_{nr}= E/(n m_h ) -1 \ll 1$ while $\varepsilon$ is now treated on the right-hand side of 
\eqref{eq:RnpN} as the integration variable.

The second point to note is that we introduced an additional term, $-\, a\, \varepsilon \, \tilde{n}^{\,p}$, in the exponent 
on the right-hand side of \eqref{eq:RnpN} which represents the sub-leading correction in $\varepsilon\ll 1$ to the 
expression in \eqref{eq:FB2}. In other words,  we have  assumed that,
\[
Q_{\tilde{n} \gg 1}(\tilde{n},\varepsilon) \,=\, +\, 0.85\, \sqrt{\tilde{n}} \, -\, a\, \varepsilon \, \tilde{n}^{\,p} \,+\, 
{\rm higher\,\,orders\,\, in\,\,} \varepsilon\,,
\label{eq:Qln2}
\]
where $a$ and $p$ are some positive constants.
The integral we have to compute is,
\[
{\cal R}_n \,\sim \,
e^{\, \frac{\tilde{n}}{\lambda}\, \left( 0.85\, \sqrt{\tilde{n}} \,+\, \log \tilde{n} \,-\, \tilde c
\right)}\,\,
\int d \varepsilon\,
e^{\, \frac{\tilde{n}}{\lambda}\, \left( \frac{3}{2} \log \varepsilon \,-\, a\, \varepsilon \, \tilde{n}^{\,p}
\right) } \,,
\label{eq:RnpNsp}
\]
where
\[
{\tilde{c}} \,=\, \log 4 \,+\, \frac{3}{2} \log 3\pi\,-\, \frac{1}{2}\,.
\]
The saddle-point dominating the integral over $\varepsilon$ is,
\[ 
\varepsilon_{\star} \,=\, \frac{3}{2}\, \frac{1}{a} \, \frac{1}{\tilde{n}^{\,p}}\,,
\label{eq:powersup}
\]
and the value of the original integral for ${\cal R}_n(\varepsilon)$ at the saddle-point is,
\[
{\cal R}_n (\varepsilon_\star) \,\sim \,
\exp \left[\frac{\tilde{n}}{\lambda}\, \left( 0.85\, \sqrt{\tilde{n}} \,+\, \left(1-\frac{3p}{2}\right) \log \tilde{n} \,-\, {\rm const}' \right)
\right]
 \,.
\label{eq:RnpNsp2}
\]
The constant term ${\rm const}' $ is,
\[
{\rm const}' \,=\, \tilde{c}\,+\, \frac{3}{2}\left(1-\log \frac{3}{2a}\right) \,=\, 
1\, +\, \log 4 \,+\, \frac{3}{2} \log 2a \pi\,.
\]
Setting $a=1$, we have ${\rm const}' \,\simeq\, 5.14$.

Now, assuming that the coefficient in front of the logarithmic term in $\varepsilon$ in \eqref{eq:RnpNsp2} is positive, 
i.e. $0< p<2/3$, the exponent is negative at small $\tilde{n}$, positive at large $\tilde{n}$ 
and crosses zero at some value $\tilde{n}_*$.
For example, for $p=1/2$ and $a=1$, which corresponds to the NLO correction 
$-\, a\, \varepsilon \, \tilde{n}^{\,p} \,=\, -\, \varepsilon \,\sqrt{\tilde{n}}$, 
the value of  $\tilde{n}_* \,\simeq 5.55$.

In the alternative scenario, where the coefficient in front of the logarithm is negative,
for example at $p=1$, the function in the exponent of  \eqref{eq:RnpNsp2} has a more complicated behaviour
with a local minimum at intermediate values of $\tilde{n}$. Nevertheless at larger $\tilde{n}$, the function
is again monotonic and crosses over from negative to positive values at $\tilde{n}_* \,\simeq 7.2$.

\medskip

It then follows that the value of $\tilde{E}_* \,=\, (1+\varepsilon_\star)\, {n}_* \,\simeq\, \tilde{n}_* \,:=\, C\,=\, {\rm const}.$
As a result, we can write the Higgsplosion scale $E_*$ as,
\[
E_*\,=\, C\, \frac{m_h}{\lambda} \,.
\label{eq:Estarfin}
\]
It is also easy to verify that this conclusion is
consistent within the validity of the non-relativistic limit.

The parametric dependence of the Higgsplosion energy $E_*$ on the particle mass and the inverse coupling constant is
reminiscent of another famous dynamically induced scale in the electroweak theory --- the mass of the 
sphaleron solution \cite{Manton:1983nd,Klinkhamer:1984di},
$M_{\rm sph}\,=\, {\rm const}\, \frac{m_W}{\alpha_w}$. Both scales are non-perturbative and semi-classical in nature. They
do not appear in the Lagrangian of the theory, but rather characterise the energy scale where the transition to novel dynamics involving multi-particle
states occurs.\footnote{In the case of sphalerons, the new dynamics is that of the non-perturbative $(B+L)$-violating
transitions between multi-particle initial and final states.} The sphaleron, however, does not occur in the pure scalar sector of the theory and requires
the $SU(2)$ gauge theory in the Higgs phase.

\medskip
\subsection{Comment on the possibility of non-analytic corrections in $\varepsilon$ to the large-$n$ WKB amplitudes}
\label{sec:logeps}
\bigskip

Before closing this section, we would like to consider the possibility that our semi-classical rate ${\cal R}_n(E)$ 
could also contain non-analytic corrections in $\varepsilon$ in the near-threshold kinematics  $\varepsilon \to 0.$

The discussion in Ref.~\cite{Khoze:2017ifq}, and the resulting realisation of Higgsplosion,
was based on the simplifying assumption that the currently unknown higher-order corrections to the 
$1^*\to n$ amplitudes (or form factors) are analytic in the parameter  
$\varepsilon$, at $\varepsilon \sim 0$. Indeed taking the leading-order term in the Taylor expansion $\sim \varepsilon^1$ of the 
correction to be of the form $-\, a\, \varepsilon \, \tilde{n}^{\,p}$, we saw in the last section that the saddle-point value of 
$\varepsilon$ was power-suppressed at $\tilde{n}=\lambda n \gg 1$, and that the rate ${\cal R}_n$ at this point was 
exponentially large  \eqref{eq:RnpN} in the large $\tilde{n}$ limit, thus signalling Higgsplosion.

\medskip

However the reader may question what would be the effect of adding to the exponent in \eqref{eq:RnpN}
a {\it non-analytic}  correction, in particular a slowly varying one, of the type $ d(\tilde{n}) /\log \varepsilon$.  
In the limit $\varepsilon \to 0$ and $\tilde{n}$ held fixed,  this correction vanishes, i.e. $ d(\tilde{n}) /\log \varepsilon \to 0$. Thus, it
does not modify the known semi-classical expression for the amplitude on the $n$-particle threshold. Since the  dependence on $\varepsilon$ 
is now inverse logarithmic rather than power-like, 
this correction does affect the $0< \varepsilon< 1$ behaviour of the rate \eqref{eq:Rnp2}
in the large $\tilde{n}$ limit which now becomes,
\[
{\cal \tilde{R}}_n(E)\,= \,  
\exp \left[ \frac{\tilde{n}}{\lambda}\, \left( \frac{3}{2}\log \frac{\varepsilon}{3\pi} \,+\,
0.85\, \sqrt{\tilde{n}} \,+\, \frac{d( \tilde{n})}{ \log \varepsilon} \,+\, \log \tilde{n}\,
\,-\, {\rm const}
\right)\right] \,.
\label{eq:RnpN2}
\]
If the coefficient function  $d( \tilde{n})$ is positive and grows with $\tilde{n}$ faster than $0.85 \sqrt{\tilde{n}}$
it would result in the exponential suppression of the rate and prevent Higgsplosion from being realised.

Can such non-analytic corrections be generated in a simple quantum theory? 
One could ask why should all corrections to the exponent of the $n$-particle rate \eqref{eq:Rnp2} be analytic,
if the term $\sim \log \varepsilon$, that was already present in the original rate \eqref{eq:Rnp2}, is not.

\medskip 

However, we know that the existing $\sim \log \varepsilon$ term in the exponent of the rate \eqref{eq:RnpN2} arises from the volume of the 
$n$-particle phase-space, and has nothing to do with the amplitudes themselves,
\[
\int d\Phi_n \,\sim \,  \int_0^{\varepsilon_{nr}} d \varepsilon\, \left(\frac{\varepsilon}{3 \pi}\right)^{\frac{3n}{2}} 
\,\sim\, 
\exp \left[ n\, \left( \frac{3}{2}\log \frac{\varepsilon}{3\pi}
\right)\right] \,.
\label{eq:RnpNgrrr}
\]
The entire logarithmic factor came from taking the logarithm of the $n$-particle non-relativistic 
phase-space. We also note that there is no dependence on the coupling constant in this factor.  Thus, this particular logarithmic term is well-understood and has nothing to do with the amplitudes. On the other hand, in order to generate
a term of the type  $ d(\tilde{n}) / \log \varepsilon$, one needs an effect coming from the amplitudes directly, since it does not arise as a correction to the 
phase-space volume. Therefore, it has to be (twice) the leading correction in $\varepsilon$ to the scattering amplitude on threshold.

We do not have a rigorous argument to prove that the non-analytic corrections of the form  $ d(\tilde{n}) / \log \varepsilon$ 
cannot be generated
in the large-$\lambda n$  amplitudes,
\[
{\cal M}_{1\to n} \,\stackrel{?}{=}\, {\cal M}_{1\to n}^{\rm threshold} \times \exp \left[-\,\frac{n \, d(\lambda n)}{2 \log 1/\varepsilon}\right]\,.
\]
However, we can offer an intuitive reasoning against the appearance of such terms. Let us consider the amplitudes ${\cal M}_{1\to n}$ near its $n$-particle 
threshold  as a function of two variables, $\lambda n$ and $\varepsilon$. Since the amplitude of interest is always near the threshold,
we will assume that $\varepsilon \ll 1$ but will allow the second argument, $\lambda n$, to vary freely from small to large values.
In the regime where $\lambda n$ is small, the ordinary perturbation theory is applicable and the amplitude is of the form,
\[
\frac{1}{n}\, \log \left({\cal M}_{1\to n}/ {\cal M}^{\rm tree}_{1\to n}\right)\,=\, \ \sum_{p=1}^\infty (\lambda n)^p \, B_p (\varepsilon)\,,
\label{eq:logpert}
\]
where the sum is over the loop orders in perturbation theory. The functions $B_p (\varepsilon)$ describe the kinematic dependence 
at each loop order, and since we are interested only in the near-threshold behaviour, we can consider their small $\varepsilon$ expansion. 
We expect that in a QFT with a mass gap and in the number of dimensions not less than 4,
all perturbative functions $B_p (\varepsilon)$ should be analytic in the vicinity of $\varepsilon \sim 0$,
so that,
\[
B_p (\varepsilon) \,=\, B^{(0)}_p\,+\, B^{(1)}_p \varepsilon \,+\, B^{(2)}_p \varepsilon^2 \,+\, \ldots\,,
\label{eq:logpert2}
\]
where the constant $B^{(0)}_1$ for the $1$-loop function is equal to the constant $B$ in \eqref{eq:FB}. 
Any appearance of the non-analytic contributions to $B_p (\varepsilon)$ of the type $1/ \log \varepsilon$ would imply
that a derivative with respect to $\varepsilon$ of the perturbative amplitude in \eqref{eq:logpert} will be singular at $\varepsilon=0$.
This is equivalent to saying that a derivative with respect to an external momentum $\vec{p}_i$
of a non-relativistic perturbative amplitude at small values of $\vec{p}_i$ would be infinite. However the Feynman diagrams 
of our scalar theory with a mass gap and in the number of dimensions not less than 4, are regular and differentiable in the 
infrared regime of  $\varepsilon \to 0$ or $p_i^2 \to 0$. Hence we will take that the behaviour \eqref{eq:logpert} and \eqref{eq:logpert2}
is justified in the regime of small $\lambda n$.

The next step is to resum the perturbation theory in $\lambda n$ for each fixed power of $\varepsilon$. 
The intuitive expectation is that while this resummation is non-trivial, it should not generate any new functional
dependence on $\varepsilon$ that has not already been there. The formally resummed result can then be continued
to the large $\lambda n$ limit, without generating the dangerous non-differentiable dependence on $\varepsilon$. 

The intuitive argument presented above is not a proof and poses an interesting problem for a future study.\footnote{In fact, it was proposed in Ref.~\cite{Gorsky:1993ix} that an additional kinematic suppression could occur non-perturbatively.
In our view their effect has more to do with the unitarity restoration, i.e. Higgspersion, and is already accounted for. But the whole point certainly deserves further study.}
This question 
of the possibility of the non-analytic and non-differentiable corrections to the amplitudes in the near-threshold limit
should ultimately be settled with an explicit calculation.
We expect that it should be possible to carry out such a calculation by introducing a small $\varepsilon$ and extending the existing semi-classical 
analysis for the amplitudes on 
the multi-particle thresholds.

\medskip
\section{Systematics of loops with Higgspersion}
\label{sec:loops}

\medskip
\subsection{Computing a loop with the propagator in the classical background}
\label{sec:proploop}
\bigskip

We will start by considering the scalar field theory \eqref{eq:nossb} with unbroken $Z_2$ symmetry, and postpone the
discussion of the broken theory \eqref{eq:ssb} to Section~\ref{sec:proploopSmith}.
As we have already explained, the generating functional for all tree-level amplitudes on $n$-particle thresholds is
given in this model by the classical solution \eqref{eq:sol_nossb}. 

\bigskip

\noindent The aim of this section is to compute the leading-order quantum corrections to these amplitudes 
in the case where a non-trivial finite Higgsplosion scale $E_*$ is present. 
The leading order calculation (in the absence of Higgsplosion) was performed in \cite{Voloshin:1992nu},
extended to the spontaneously broken theory in \cite{Smith:1992rq}, and generalised 
in \cite{Libanov:1994ug} to include 
all higher-loop effects by exponentiation to the leading order in $\lambda n$.

We begin by following closely the original leading-loop calculation of Voloshin in \cite{Voloshin:1992nu}, and then
explain how it should be modified to reflect the appearance of the Higgsplosion scale $E_*$. This will allow us to assess the effect of Higgsplosion on the RG running of the parameters of the theory, including their asymptotic safety.
We will also see that the so-called finite terms arising from the quantum effect are the same as those computed in
 \cite{Voloshin:1992nu,Smith:1992rq,Libanov:1994ug} up to corrections of the order ${\cal O}(m^2/E_*^2)$. 
 These considerations will pave the way for computing precision observables in Higgsplosion in Section~\ref{sec:qft}.

\bigskip

The quantum corrections to the tree-level amplitudes \eqref{eq:expsc12} are obtained by expanding 
around the classical field, $\phi(x) = \phi_0(x) + \phi_q(x)$,
so that the Euclidean Lagrangian \eqref{eq:nossb} for the quantum fluctuation $\phi_q$ becomes,
\[
{\cal L} \,=\, \frac{1}{2} (\partial_\mu \phi_q)^2 \,+\,  \frac{1}{2} \left(m^2+3\lambda \phi_0^2\right) \phi_q^2 \,+\, 
\lambda \phi_0 \phi_q^3 \,+\, \frac{\lambda}{4} \phi_q^4\,.
\label{eq:Lnossb_phiq}
\]
One then integrates out $\phi_q(x)$
using the background field perturbation theory. 

It follows that the generating functional of the amplitudes 
in the full quantum theory is obtained by promoting the classical solution $\phi_0$ into the quantum
expectation value
$\langle \phi \rangle \,=\, \phi_0 + \langle \phi_q \rangle$. Individual amplitudes are then computed via
\[
 \langle n|\phi |0\rangle\,=\, 
\left(\frac{\partial}{\partial z_0}\right)^{n} \,  (\phi_0 +  \langle \phi_q \rangle) \, |_{z_0=0} \,.
\label{eq:Aquant}
\]
This provides the generalisation to full quantum theory \cite{Voloshin:1992nu,Libanov:1994ug}
of the tree-level formalism of Brown \cite{Brown:1992ay} for computing $1^* \to n$ amplitudes on 
$n$-particle mass-thresholds.

The matrix element $ \langle \phi_q \rangle$ is computed using the Feynman rules following from the 
action \eqref{eq:Lnossb_phiq}. It is easy to see that the one-loop contribution to $ \langle \phi_q \rangle$
comes from the tadpole diagram, which contains the three-point vertex from \eqref{eq:Lnossb_phiq} 
with two attached propagators: one external, $G(y,x)$, and one forming the loop, $G(x,x)$,
\[
\langle \phi_q (y) \rangle_{\rm 1-loop}\,=\,
(-3\lambda) \int d^4 x \, G(y,x) \, \phi_0(x) \, G(x,x)\,,
\label{eq:phi1_loop}
\]
where $G(x_1,x_2)$ is the propagator for the scalar field $\phi_q$ in the background of the classical solution,\footnote{To distinguish
the propagator in the background of $\phi_0(t)$ from the propagator in the trivial background, we call it
$G$ rather than $\Delta$. We also continue working in Euclidean space and thus drop the $T$-ordering in the propagator.}
\[ 
G(x_1,x_2)\,=\,  \langle 0|\phi_q(x_1) \phi_q(x_2)|0\rangle\,,
\]
which satisfies the equation,
\[
\left(-\big(\dfrac{\partial}{\partial x_1}\big)^2 \,+\,  m^2\,+\,3\lambda \,\phi_0(x_1)^2\right)\, G(x_1,x_2) \,=\, \delta^{(4)} (x_1-x_2)\,.
\label{eq:eq_prop}
\]
The leading-order quantum correction $\langle \phi_q \rangle_{\rm 1-loop}$ obtained via \eqref{eq:phi1_loop}
is the solution of the differential equation,
\[
\left(-\big(\dfrac{\partial}{\partial x}\big)^2 \,+\,  m^2\,+\,3\lambda \,\phi_0(x)^2\right)\, \langle \phi_q (x) \rangle_{\rm 1-loop} \,=\, 
-\, 3 \lambda \, \phi_0(x) \, G(x,x)\,.
\label{eq:decorr}
\]
This equation is derived by acting with the differential operator appearing on the left, 
on both sides of \eqref{eq:phi1_loop} and using the definition of the propagator $G$ in \eqref{eq:eq_prop}.

It will also be useful for our purposes to write down the equation for the quantum corrected 
generating function $\phi=\phi_0 + \langle \phi_q \rangle_{\rm 1-loop}$. Using the fact that $\phi_0$ satisfies the classical
equation, $-\partial^2 \phi_0+m^2 \phi_0 +\lambda \phi_0^3 =0$, and that the quantum correction
satisfies Eq.~\eqref{eq:decorr}, 
it follows from combining the two, that the full generating function $\phi(x)$ is the solution to
\[
-\partial^2 \phi(x) \,+\, m^2 \phi(x) \,+\,\lambda \phi(x)^3\,+\, 3\lambda \,\phi_0(x)\, G(x,x) \,=\,0\,.
\label{eq:fullphi}
\]

\bigskip

We will ultimately need to compute $G(x,x)$ at  coincident points,
but for now we consider the general case of $x_1\neq x_2$.
It is convenient to use the mixed coordinate-momentum representation $G_\omega (\tau_1,\tau_2)$ for the propagator, 
i.e. perform the 3D Fourier transform to the 3-momentum, but keep the Euclidean time coordinate $\tau$,
\[
G(x_1,x_2) \,=\, \int \frac{d^3 p}{(2\pi)^3}\, e^{i \vec p (\vec{x}_1 - \vec{x}_2)}\, G_\omega (\tau_1,\tau_2)\,.
\label{eq:prop_10}
\]
The partial differential equation \eqref{eq:eq_prop} then becomes an ordinary second order differential equation,
\[
\left(-\dfrac{d^2}{d \tau_1^2} \,+\,  \omega^2\,+\,3\lambda \,\phi_0(\tau_1)^2\right)\, G_\omega(\tau_1,\tau_2) \,=\, 
\delta (\tau_1-\tau_2)\,,
\label{eq:eq_prop2}
\]
where once again,
\[ 
\omega \,:=\, \omega_p \,=\, \sqrt{\vec{p}^{\,2} +m^2}\,.
\]

The classical solution \eqref{eq:sol_nossb} entering the equation \eqref{eq:eq_prop2} 
should be Wick rotated to the Euclidean time $\tau$. To simplify the resulting expressions, 
we also introduce a constant shift as in \cite{Voloshin:1992nu},
\[
\tau\,=\, i t \, +\, \frac{1}{m} \log \frac{\lambda z_0}{8m^2} \,-\, \frac{i \pi}{2m}\,,
\]
which gives,
\[
\phi_0 (\tau)\,=\, i\,\sqrt{\dfrac{2}{\lambda}} \, \dfrac{1}{\cosh (\tau)}\,,
\label{eq:sol_nossb2}
\]

Thus, to determine the propagator in the classical background \eqref{eq:sol_nossb2} 
one should solve the inhomogeneous second-order linear ODE,
 \[
\left(-\dfrac{d^2}{d \tau_1^2} \,+\,  \omega^2\,- \,\dfrac{6}{\cosh^2(\tau_1)}\right)\, G_\omega(\tau_1,\tau_2) \,=\, 
\delta (\tau_1-\tau_2)\,.
\label{eq:eq_prop3}
\]
This is a textbook problem. The required solution of the inhomogeneous equation is obtained from the two
solutions $f_1(\tau)$ and $f_2(\tau)$ to the corresponding homogeneous equation. The result is,
\[
G_\omega (\tau_1,\tau_2) \,=\, \dfrac{1}{W}[\Theta(\tau_1-\tau_2)f_1(\tau_1)f_2(\tau_2)+\Theta(\tau_2-\tau_1)f_2(\tau_1)f_1(\tau_2)]\,,
\]
where $\Theta$'s are the step functions and $W$ is the Wronskian computed for the two homogeneous solutions,
\[
W \,=\, \det\begin{pmatrix}
f_1 & f_2\\
f_1' & f_2'
\end{pmatrix}
\,.
\]
The solutions of the homogeneous equation 
 \[
\left(-\left(\dfrac{d}{d \tau}\right)^2 \,+\,  \omega^2\,- \,\dfrac{6}{\cosh^2(\tau)}\right)\, f_{1,2}(\tau) \,=\, 
0\,,
\label{eq:eq_f12}
\]
are known \cite{Voloshin:1992nu} 
thanks to the identification of Eq.~\eqref{eq:eq_f12} as the Scr{\"o}dinger equation with an exactly solvable potential.
The first solution (which is regular at $\tau \to +\infty$) can be written in the form, 
\[
f_1(\tau) \,=\, \left(\omega^2\,+\, 3 \omega \dfrac{(e^{2\tau}-1)}{e^{2\tau}+1}\,+\, 2\,-\, \dfrac{12 \, e^{2\tau}}{(e^{2\tau}+1)^2}
\right)\, e^{-\omega \tau} \,, \quad {\rm where}\quad
u\,=\, e^\tau\,,
\]
where here and below we have temporarily set $m=1$ 
to reduce the clutter. The second solution (which is regular at $\tau \to -\infty$) is obtained from $f_1(\tau)$ by the reflection of its argument,
$f_2 (\tau) \,=\, f_1 (-\tau)$.
These expressions can also be checked by a direct substitution into \eqref{eq:eq_f12}. 
Finally, the Wronskian computed for these two solutions is given by \cite{Voloshin:1992nu},
\[
W\,=\, 2 \omega (\omega^2-1)(\omega^2-4)\,,
\]
and is independent of the $\tau$ variable (as indeed should be the case for the second order linear ODE). 
Once again we remind the reader that the mass in the expressions above was rescaled to $m=1$ 
(so that e.g. $\omega=\sqrt{\vec{p}^{\,2}+1}$ and is dimensionless), and that it can be easily recovered, when necessary, by dimensional analysis.

\medskip

To find the expression for the coincident propagator 
$G(x, x)$ in \eqref{eq:phi1_loop}, we must first evaluate the integral in \eqref{eq:prop_10} for $G(x+\Delta x, x)$ , and then take the limit as $\Delta x \to 0$ to close the loop.
However, we have to be careful with the order of limits. In fact, the momentum space integration produces a UV-finite result
only at non-zero separation between the propagator end points $x_1$ and $x_2$. When $\Delta x$ does go to zero, the 
integral $\int d^3p $ contains quadratically and logarithmically divergent terms and needs to be regulated. 
The regulated divergences are then absorbed into the renormalised parameters of the theory --- the mass squared term and the coupling
constant --- after which the UV regulator can be removed. This was the procedure adopted in \cite{Voloshin:1992nu}.
 
Since our main aim is to account for the effects of the Higgsplosion scale $E_*$, which renders the loop integrals 
over Euclidean 4-momenta finite, as in \eqref{eq:propLoop}, we must now deviate from \cite{Voloshin:1992nu}.
Our approach will be to keep a non-vanishing separation between the space-time points, so that our integrals remain finite, 
before re-writing them as the integrals over $d^4 p$. For these integrals we can implement the Higgsplosion 
cut-off $p^2\le E_*^2$ directly, as in \eqref{eq:propEX}, and then safely take limit $\Delta x\to 0$ as in \eqref{eq:propLoop}.

\medskip

To follow the approach outlined above, we expand the integrand in \eqref{eq:prop_10}
 in powers of $1/\omega$, to isolate the most sensitive terms in the UV, as $p \sim \omega \to \infty$.
 With no loss of generality we can present \eqref{eq:prop_10} as follows:
\[
G(\tau+\Delta \tau, \tau)\,=\,
\int \dfrac{d^3 p}{(2\pi)^3} \,\, e^{-\omega \Delta \tau}\,\frac{1}{2 \omega}\, \left( 1 \,+\,
\frac{1}{ \omega} \, A \,+\,
\frac{1}{\omega^2} \, B  \,+\,  {\cal O}\left(\frac{1}{\omega^3}\right)\right)\,.
\label{eq:Gominv}
\]
Here we have set the non-zero separation between the space-time points to be along the time direction and kept $\Delta \tau$ positive.
The factors $A$ and $B$ appearing on the right-hand side of \eqref{eq:Gominv} are $\omega$-independent functions of 
$\tau$ and $\Delta \tau$. Using the formulae for $f_1(\tau)$, $f_2(\tau)$ and the Wronskian we find,
\begin{eqnarray}
A &=&\dfrac{12 \, e^{2\tau}}{(e^{2\tau}+1)^2}\, \Delta \tau  \,+\, {\cal O}( \Delta \tau^2)
\,=\, -\dfrac{3 \lambda}{2}\, \phi_0 (\tau)^2 \, \Delta \tau  \,+\, {\cal O}( \Delta \tau^2)\,,
\\
B &=&\dfrac{12 \, e^{2\tau}}{(e^{2\tau}+1)^2}  \,+\, {\cal O}( \Delta \tau)
\,=\, -\dfrac{3 \lambda}{2}\, \phi_0 (\tau)^2  \,+\, {\cal O}( \Delta \tau)\,.
\end{eqnarray}
To obtain the simple form for $A$ and $B$ used above, we anticipate the ultimate $\Delta \tau\to 0$ limit and Taylor-expand the expressions in $A$ and $B$ to the first non-vanishing order in  $\Delta \tau$.
Note however, that while we can treat 
$\Delta \tau$ as a small parameter, the combination $w \Delta \tau$ is not small at high values of momenta
where the exponential factor $e^{-\omega \Delta \tau}$ regulates the integrand in \eqref{eq:Gominv}. This factor is therefore left unchanged.

After integrating over $d^3 p$, the first three terms in the brackets on the right-hand side of \eqref{eq:Gominv} give rise respectively to the quadratic, linear, and logarithmic dependence on either the $1/\Delta \tau$ parameter, or on the UV cut-off 
momentum scale $\Lambda_{UV}$, depending on whether $\Lambda_{UV} \Delta \tau$ is $\gg 1$ or $\ll 1$.
The fourth and final term in \eqref{eq:Gominv} goes as $\int dw/w^2$ and is regular in the UV  at $\Delta \tau=0$.
It will be useful to represent  \eqref{eq:Gominv} as follows,
\[
G(\tau+\Delta \tau, \tau) \,=\, G_{\rm I} \,+\, G_{\rm II}
\,+\, G_{\rm III} \,+\, G_{\rm finite}
\,.
\label{eq:Gominv4}
\]

\noindent
We can now evaluate the  integrals $G_{\rm I}$, $G_{\rm II}$, $G_{\rm III}$ and $G_{\rm finite}$.
For the first integral, we have,
\[
G_{\rm I}(\tau+\Delta \tau, \tau) \,=\,
\int \dfrac{d^3 p}{(2\pi)^3} \,\frac{1}{2 \omega}\, e^{-\omega \Delta \tau}
 \,=\,  \int \frac{d^4 p}{(2\pi)^4}\,\frac{1}{p^2 +m^2 }\, e^{i p_0 \Delta \tau}\,,
\label{eq:prop1st} 
\]
where we have used the familiar relation between the 4D and the 3D integrals for the
free propagator,
\[
\Delta_0(\tau+\Delta \tau, \tau) \,:=\, 
 \int \frac{d^4 p}{(2\pi)^4}\,\frac{1}{p^2 +m^2 }\, e^{i p_0 \Delta \tau}
 \,=\, \int \dfrac{d^3 p}{(2\pi)^3} \,\frac{1}{2 \omega}\, e^{-\omega \Delta \tau}\,,
\label{eq:prop3dtau}
\]
which is based on the use of the residue theorem.

For non-vanishing $\Delta \tau$ the integral is finite and does not require any UV cut-off.
The expression \eqref{eq:prop1st} was derived in a theory without Higgsplosion, or in the limit of infinite Higgsplosion energy scale
$E_*$. But since the expression on the right is the ordinary free propagator in a trivial background,
we know how to modify this expression to account for the Higgsploding self-energy. Higgsplosion is introduced  
exactly as in \eqref{eq:propEX}, and now we can take the limit $\Delta \tau \to 0 $ 
and find the contribution of the first term in \eqref{eq:Gominv}
to the closed loop. It is given by,
\[
G_{\rm I}(x, x) \,=\,  \int_{p^2\le E_*^2} \, \frac{d^4 p}{(2\pi)^4}\,\,\frac{1}{p^2 +m^2 }
\,=\,  \dfrac{1}{16\pi^2} \, 
\left(E_*^2 \,-\, m^2\, \log \dfrac{E_*^2+m^2}{m^2}\right)
\,.
\label{eq:GI} 
\]

\noindent To evaluate $G_{\rm II}$ and $G_{\rm III}$ 
we note the useful identity obtained by differentiating $G_{\rm I}(\tau+\Delta \tau, \tau)$ on the first line in \eqref{eq:prop1st} with respect to $m^2$, 
\[
\dfrac{\partial}{\partial m^2}\,  \int \dfrac{d^3 p}{(2\pi)^3} \,\frac{1}{2 \omega}\, e^{-\omega \Delta \tau}
\,=\,-\,\dfrac{1}{2}\,\int \dfrac{d^3 p}{(2\pi)^3} \,\left(\frac{1}{2 \omega^2}+ \frac{\Delta \tau}{2 \omega^3} \right) e^{-\omega \Delta \tau}\,,
\]
and recognising the expression on the right-hand side as precisely the sum of the integrals appearing in $G_{\rm II}$ and $G_{\rm III}$.
Since $B=12 \, e^{2\tau}/(e^{2\tau}+1)^2$ and $A=B \Delta \tau$, we find that
\[
G_{\rm II}\,+\,G_{\rm III}\,=\, -\,\dfrac{24 \, e^{2\tau}}{(e^{2\tau}+1)^2}\, \dfrac{\partial}{\partial m^2}\,G_{\rm I}\,.
\label{eq:3plus4}
\]
Differentiating our result for $G_{\rm I}$ on the right-hand side of \eqref{eq:GI}, we find the closed-form expression for 
the sum of $G_{\rm II}$ and $G_{\rm III}$ at $\Delta \tau=0$,
\[
G_{\rm II}(x,x)\,+\,G_{\rm III}(x,x)\,=\, 
\dfrac{1 }{2\pi^2} \, \dfrac{3\,e^{2\tau}}{(e^{2\tau}+1)^2}\, 
\left(\log \dfrac{E_*^2+m^2}{m^2} \,-\, \dfrac{E_*^2}{E_*^2+m^2}\right)\,.
\label{eq:3plus4_res}
\]

The final term contributing to the propagator in \eqref{eq:Gominv4}, $G_{\rm finite}$,
contains no UV-sensitive contributions. It can be computed directly at $\Delta x=0$ and 
one can set $E_*\to \infty$ without the need of introducing any UV cut-off.
In this case we find the same expression as in \cite{Voloshin:1992nu},
\[
G_{\rm finite}(x,x)\,=\, \dfrac{1 }{2\pi^2} \, \dfrac{6\,e^{2\tau}}{(e^{2\tau}+1)^2}\,-\,
  \dfrac{6\,e^{4\tau}}{(e^{2\tau}+1)^4}\, F\,,
\label{eq:Gfin}
\]
where 
\[
F\,=\, \dfrac{\sqrt{3}}{2\pi^2} \left(\log\dfrac{2+\sqrt{3}}{2-\sqrt{3}}\,-\, i \pi \right)\,.
\label{eq:Fnossb}
\]

Collecting the expressions for the individual contributions in equations 
\eqref{eq:GI}, \eqref{eq:3plus4_res}-\eqref{eq:Gfin} 
we arrive at the final result for the propagator loop in the background field of the unbroken scalar theory with the
Higgsplosion energy scale $E_*$.
It can be slightly simplified in the limit $m^2 / E_*^2 \ll 1$, where we get,
\[
 G(x,x)\,=\, \dfrac{1}{16\pi^2} \, 
\left(E_*^2 \,-\, m^2\, \log \dfrac{E_*^2}{m^2}\right)\,+\, 
\dfrac{1 }{2\pi^2} \, \dfrac{3\,e^{2\tau}}{(e^{2\tau}+1)^2}\, 
\left(\log \dfrac{E_*^2}{m^2} \,+\, 1\right)
\,-\,
  \dfrac{6\,e^{4\tau}}{(e^{2\tau}+1)^4}\, F,
  \label{Gxx1}
\]
or expressing everything in terms of the Brown's solution $\phi_0$, we can write an
alternative equivalent representation,
\[
 G(x,x)\,=\, \dfrac{1}{16\pi^2} \, 
\left(E_*^2 \,-\, m^2\, \log \dfrac{E_*^2}{m^2}\right)\,-\, 
\dfrac{1 }{2\pi^2} \, \dfrac{3\,\lambda\, \phi_0^2}{8}\, 
\left(\log \dfrac{E_*^2}{m^2} \,+\, 1\right)
\,-\,
  \dfrac{3\,\lambda^2\, \phi_0^4}{32}\, F\,.
  \label{Gxx2}
\] 

\medskip

Recalling the defining equation \eqref{eq:fullphi} for the quantum-corrected generating functional $\phi(x)$,
makes it obvious that there is a naturally occurring combination,
\[
m^2 \phi_0(x) \,+\,\lambda \phi_0(x)^3\,+\, 3\lambda \,\phi_0(x)\, G(x,x) \,.
\]
It makes sense to re-absorb the first two terms on the right-hand side of \eqref{Gxx2} into the 
definitions of the renormalised coupling parameters of the theory in the following way,
\begin{eqnarray}
\bar\lambda_* &=& \lambda\,-\, \frac{9\,\lambda^2}{8 \pi^2} \left(\log \frac{E_*}{m} +\frac{1}{2} \right)\,,
\label{eq:lamR}\\
\bar{m}^2_* &=& m^2 \,+\, \frac{3\,\lambda}{16 \pi^2}
\left(E_*^2 \,-\, m^2\, \log \dfrac{E_*^2}{m^2}\right)\,,
\label{eq:m2R}
\end{eqnarray}
where the subscript$_*$ indicates that these renormalised parameters correspond to the theory with the
Higgsplosion scale $E_*$.
Eqs.~\eqref{eq:lamR} and \eqref{eq:m2R}
define a {\it finite} renormalisation of the parameters, since $E_*$ is a fixed finite non-perturbative scale 
of the theory; it is neither the RG scale nor the UV cut-off scale. 
Such finite renormalisation of the model parameters is what we would expect in the theory with Higgsplosion: 
such a theory does not contain UV divergences and flows to an asymptotically safe theory, where 
all couplings approach constant values in the UV, as was first explained in \cite{Khoze:2017lft}.
Our new results,  Eqs.~\eqref{eq:lamR}-\eqref{eq:m2R}, provide a precise definition of these renormalised parameters 
in the UV fixed point. In Section~\ref{sec:run} we will interpret these equations in terms of the running parameters and
show they have the standard running with the correct slopes at low values of the RG scale, $\mu <E_*$, and
flatten out and approach fixed points above the Higgsplosion scale $\mu > E_*$.

\medskip 

Finally, we are now in a position to compute the leading-order quantum correction 
$\langle \phi_q (x) \rangle_{\rm 1-loop}$ to the generating function
of $n$-point amplitudes on multi-particle mass thresholds.
The defining equation for this quantity is Eq.~\eqref{eq:decorr}, and since we have already computed the
propagator loop $G(x,x)$, all the quantities in this equation are known.
In terms of the renormalised parameters \eqref{eq:lamR} and \eqref{eq:m2R}, Eq.~\eqref{eq:decorr} 
(with $\bar{m}$ set to $1$) reads \cite{Voloshin:1992nu},
\[
\left(-\dfrac{d^2}{d\tau^2} \,+\,  1\,-\, \dfrac{24\,e^{2\tau}}{(e^{2\tau}+1)^2}
\right)\, \langle \phi_q (\tau) \rangle_{\rm 1-loop} \,=\, 
i\, 18\lambda \frac{\sqrt{8}}{\sqrt{\lambda}} \, F\, \dfrac{e^{5\tau}}{(e^{2\tau}+1)^5}\,.
\label{eq:decorr2}
\]
The solution is given by the following expression \cite{Voloshin:1992nu},
\[
\langle \phi_q (\tau) \rangle_{\rm 1-loop} \,=\, 
-i\, \frac{3\lambda}{4}\, \frac{\sqrt{8}}{\sqrt{\lambda}} \, F\, \dfrac{e^{5\tau}}{(e^{2\tau}+1)^3}\,,
\label{eq:solphi1}
\]
which can be checked by substitution. Adding this quantum correction to the tree-level contribution $\phi_0(\tau)$,
we derive the on-threshold amplitudes, cf. Eq.~\eqref{eq:expsc12},
\[
 \langle n|\phi |0\rangle \,=\, 
\left(\frac{\partial}{\partial z_0}\right)^{n} \phi \, |_{z_0=0} \,=\,n! \left(\frac{\bar\lambda_*}{8\bar{m}_*^2}\right)^{(n-1)/2}
\left(1\,-\, \bar\lambda_* (n-1)(n-3)\, \frac{F}{16} \right)\,.
\label{eq:A1l}
\]
This is the same expression as that derived in \cite{Voloshin:1992nu}. Where did the dependence on the Higgsplosion scale
$E_*$ go? First, it enters the definition of the renormalised coupling and mass parameters given by Eqs.~\eqref{eq:lamR}-\eqref{eq:m2R}.
Secondly, there exists an additional effect, which is the correction to the finite part $F$ itself. Specifically,
$F$ in \eqref{eq:A1l} is given by the expression in \eqref{eq:Fnossb} plus the corrections of the order $\sim m^2/E_*^2$ which we have 
omitted in deriving the expression in \eqref{eq:Fnossb}. Section~\ref{sec:qft} is specifically dedicated
to computations of finite corrections from the Higgsplosion scale to precision observables.

\medskip
\subsection{Running couplings}
\label{sec:run}
\bigskip
In the absence of Higgsplosion, i.e. in the limit where the Higgsplosion scale is above some regulating UV cut-off, $\Lambda_{UV} < E_{*}\to\infty$, the divergent terms are absorbed into the renormalised parameters as follows,
\begin{eqnarray}
\bar\lambda &=& \lambda\,-\, \frac{9\,\lambda^2}{8 \pi^2} \left(\log \frac{\Lambda_{UV}}{m} +\frac{1}{2} \right)\,,
\label{eq:lamRUV}\\
\bar{m}^2 &=& m^2 \,+\, \frac{3\,\lambda}{16 \pi^2}
\left(\Lambda_{UV}^2 \,-\, m^2\, \log \dfrac{\Lambda_{UV}^2}{m^2}\right) \,.
\label{eq:m2RUV}
\end{eqnarray}
These equations are fully analogous to Eqs.~\eqref{eq:lamR}-\eqref{eq:m2R} computed in the presence of
Higgsplosion, except the role of $E_*$ is now played by the UV cut-off $\Lambda_{UV}$, which is interpreted as the cut-off on the 4-momentum integration $\int d^4p$.

\begin{figure}[t]
\begin{center}
\includegraphics[width=0.5\textwidth]{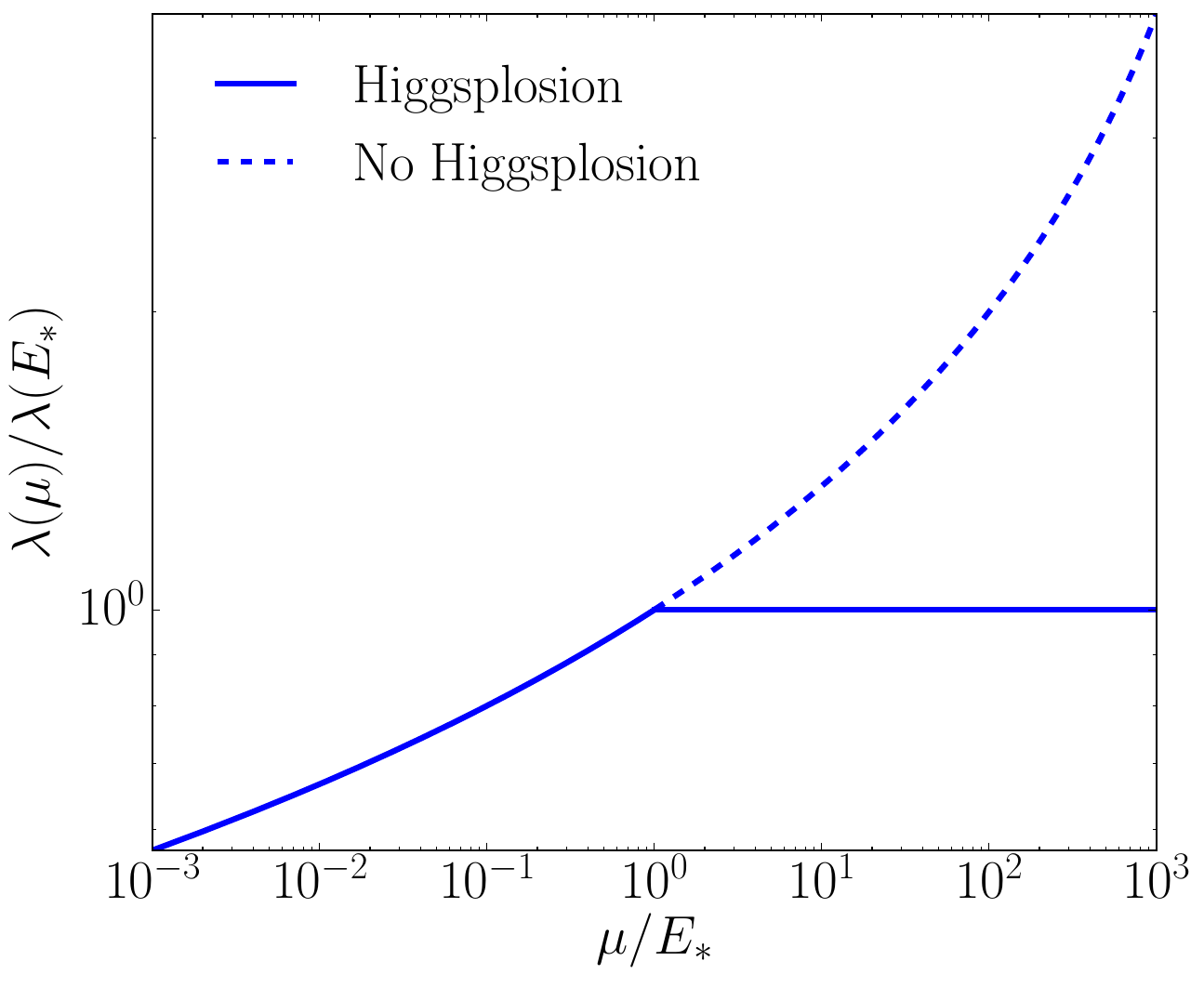}
\end{center}
\caption{Running coupling $\lambda(\mu)$ in the $\phi^4$ model with Higgsplosion (solid line). 
The profile of $\lambda(\mu)$ of the same model in the absence of the Higgsplosion scale is shown as the dashed line.
The slope of $\lambda(\mu)$ in the presence of Higgsplosion becomes flat for $\mu>E_*$ where
the coupling freezes at the constant value $\bar{\lambda}_*$. This is the UV fixed point of the asymptotically safe theory.}
\label{fig:lamrun}
\end{figure}

These expressions for the renormalised couplings have a familiar interpretation in terms of the running couplings at the RG scale $\mu$.
In the Wilsonian approach to renormalisation, one begins with the theory defined at some UV cut-off scale $\Lambda_{UV}$, where the bare parameters of the theory are specified.
The UV cut-off is then lowered from $\Lambda_{UV}$ to $\mu$ by 
integrating out all degrees of freedom in the momentum shell $\mu^2 < p^2 < \Lambda_{UV}^2$. 
The renormalised couplings $\lambda$ and $m^2$ subsequently evolve to new values, specific to this RG scale $\mu$. 
Hence in order to infer the running couplings $\lambda(\mu)$ and $m^2(\mu)$ from the one-loop computation described above, one has to restrict the integration over the loop momenta to the 
interval $[\mu, \Lambda_{UV}]$. Adapting the expression for $\bar{\lambda}$ accordingly, ignoring the constant terms and 
assuming $\mu \gg m$, we find,
\[
1/\lambda(\mu)\,=\,1/\lambda(\Lambda_{UV})\, + \,\frac{9}{8\pi^2}\int_{\mu}^{\Lambda_{UV}} \frac{d p}{p}\,.
\]
This gives the well-established one-loop solution for the running coupling in $(\lambda/4)\phi^4$ theory
\[
\lambda(\mu)\,=\, \frac{1}{\beta_0 \, \log\left(\frac{\Lambda_{\rm LP}}{\mu}\right)}\, \quad, \quad 
\beta_0= \frac{9}{8\pi^2}\,.
\]
We recognise $\beta_0$ as the one-loop coefficient of the $\beta$ function,
$\beta(\lambda)=  \frac{9}{8\pi^2} \,\lambda^2 \,+\, {\cal O}(\lambda^3)$, and the scale 
$\Lambda_{\rm LP}$ as the Landau pole of the $\phi^4$ theory without Higgsplosion.
Now consider the case where $\Lambda_{UV}$ is greater than the Higgsplosion scale so that $E_*<\Lambda_{UV}\le\Lambda_{LP}$.
When $\mu$ exceeds $E_*$, we enter the Higgsploding regime and the integral contributing to $\lambda(\mu)$ ceases to grow. 
The $\beta$ function vanishes and the coupling is frozen at $\lambda(\mu=E_*) = \bar\lambda_*$, 
the UV fixed point value given in \eqref{eq:lamR}.
The resulting running of $\lambda(\mu)$ is shown in Fig. \ref{fig:lamrun}.

The same logic and assumptions can be applied to the running of the mass. In the absence of Higgsplosion we find,
\[
m^2(\mu) \,=\, m^2(\Lambda_{UV})-\dfrac{3\lambda(\mu)}{16\pi^2}\left(\Lambda_{UV}^2-\mu^2 \,-\,  m^2 \log \dfrac{\Lambda_{UV}^2}{\mu^2}\right)\,.
\]
The expression on the right-hand side contains, as expected, quadratic and logarithmic terms depending on the UV cut-off. 
The quadratic terms in particular, introduce the high degree of fine tuning between the bare mass $m^2(\Lambda_{UV})$ and the 
radiative corrections. This is the Hierarchy problem intrinsic to QFTs with microscopic scalars. 
Once again, the effect of Higgsplosion is to freeze the RG evolution at scales above $E_*$. As a result, only the
so-called  `little hierarchy' problem remains at the scale $E_* \ll M_{\rm Pl}$.

An even more powerful simplification comes from effect of Higgsplosion on the radiative corrections to the light scalar mass, $m^2$,
 arising from the loop of heavy degrees of freedom $M \gg E_* \gg m$. In this case \cite{Khoze:2017tjt},  
the radiative correction to the bare $m^2$ parameter is given by (cf. \eqref{eq:GI}),
\[
\Delta m^2 \,=\, -3\, \lambda\,  \int_{p^2\le E_*^2} \, \frac{d^4 p}{(2\pi)^4}\,\,\frac{1}{p^2 +M^2 }
\,\simeq\,  \dfrac{-3\lambda}{16\pi^2} \, \dfrac{E_*^2}{M^2} \, E_*^2\ ,  \quad {\rm where}\, \quad  
\dfrac{E_*^2}{M^2}\, \ll 1\,.
\label{eq:GIM} 
\]

Returning to the model with a single degree of freedom, we are now in a position to understand the effect of Higgsplosion on the end result of Section~\ref{sec:proploop}. The on-threshold $1^* \to n$ amplitude given in Eq.~\eqref{eq:A1l}, now expressed in terms of the running parameters, reads,
\[
 \langle n|\phi |0\rangle \,=\, 
\left(\frac{\partial}{\partial z_0}\right)^n \phi \, |_{z_0=0} \,=\,n! \left(\frac{\lambda(\mu)}{8m^2(\mu)}\right)^{\frac{n-1}{2}}
\left(1\,-\, \lambda(\mu) (n-1)(n-3)\, \frac{F}{16} \right)\,,
\label{eq:largen}
\]
with the running parameters $\lambda(\mu)$ and $m^2(\mu)$ approaching constant UV fixed values at values of $\mu \ge E_*$.

Any corrections to $F$ due to Higgsplosion are of order $\sim$ $m^2/E^2_*$ and are ignored in the equations above. The leading effect of Higgsplosion enters through the modified running of the mass and coupling. 
We therefore expect the Higgsplosion result to match the result in \cite{Voloshin:1992nu} at scales below $E_*$. As $\mu$ exceeds $E_*$, the Higgsplosion result begins to deviate from~\cite{Voloshin:1992nu} due to the frozen couplings.

\medskip
\subsection{Spontaneously broken theory}
\label{sec:proploopSmith}
\bigskip

The approach in Section~\ref{sec:proploop} can also be directly applied to 
the broken theory, where field $\phi$ acquires a VEV.
Voloshin's calculation \cite{Voloshin:1992nu} for the model shown in Eq.~\eqref{eq:ssb}
was extended to the broken theory by Smith \cite{Smith:1992rq}. The new physical scalar boson $h=\phi-\langle\phi\rangle=\phi-v$ has mass $m_h=\sqrt{2\lambda} v$ and a new 3-point coupling. Keeping only leading order shifts and assuming $\mu\gg m_h$, the renormalised parameters are given by,
\begin{eqnarray}
\bar\lambda &=& \lambda\,-\, \frac{9\,\lambda^2}{32 \pi^2}\log \frac{\Lambda_{UV}}{m_h} \,,
\label{eq:brokenlamRUV}\\
\bar{m}_h^2 &=& m_h^2 \,-\, \frac{3\,\lambda}{8 \pi^2}
\,\Lambda_{UV}^2 \,.
\label{eq:brokenM2RUV}
\end{eqnarray}
Though some pre-factors and signs differ from the unbroken case, the degrees of divergence remain the same.
The $1^*\to n$ threshold amplitude with one-loop correction can be written as,
\[
\langle n|\phi|0\rangle \,=\, 
n!\,\left(\dfrac{\lambda(\mu)}{2m_h^2(\mu)}\right)^{(n-1)/2}\left(\ 1+ \lambda(\mu)\,n(n-1)B \right)\,,
\]
where 
\[ 
B\,=\, \dfrac{\sqrt{3}}{8\pi}\,,
\label{eq:finB}
\]
is the constant term in the broken theory, analogous to the constant $-F/16$ in the unbroken theory (see Eqs.~\eqref{eq:A1l} and \eqref{eq:Fnossb}).
Importantly, and in contrast to the unbroken case, the finite term $B$ is real and the correction is positive, 
as discussed by Smith \cite{Smith:1992rq}. As before, the leading effect of Higgsplosion is through the modified running couplings rather than the minor corrections to $B$.

\medskip
\subsection{Exponentiation of the one-loop corrections}
\label{sec:exp}
\bigskip

It is worth noting that the exponentiation of the finite terms in the one-loop corrections 
to the threshold amplitudes, as derived in Ref.~\cite{Libanov:1994ug}, will not be modified by the inclusion of Higgsplosion. The argument used in Ref.~\cite{Libanov:1994ug} 
concerns only the so-called finite part of the quantum correction, and not the UV-sensitive terms where the effect of Higgsplosion is manifest. Hence the entire construction presented in \cite{Libanov:1994ug} 
applies equally well to Higgsploding theories. The result for the exponentiated one-loop correction in the broken and unbroken theories 
respectively is,
\[
\langle n|\phi |0\rangle \,=\, 
\begin{cases}
\,\, \langle n|\phi |0\rangle_{\rm tree} \times \exp( +\, n^2 \lambda \, B)
 & :\,\,{\rm broken\, theory}\\
\,\, \langle n|\phi |0\rangle_{\rm tree} \times \exp(-\, n^2 \lambda \, F/16)
 & :\,\,{\rm unbroken\, theory}
\end{cases}\,.
\label{eq:B_unB}
\]

\medskip
\section{Precision observables in the presence of Higgsplosion}
\label{sec:qft}

Many processes in the SM occur at tree level, such as the Drell-Yan production of lepton pairs. These contain no unconstrained momenta, and thus no loop integration is required to calculate them. Only once higher-order perturbative corrections are included, do loop integrals begin to appear. Instead, we will focus on a class of observables which have no tree-level contribution. These loop-induced processes will be modified by Higgsplosion at leading order, so they provide a firm testing ground for investigating Higgsplosion effects.

\medskip
\subsection{Propagators for fermions and massive vector bosons}
\label{sec:propFV}
\bigskip

Expressions for Dyson-resummed propagators  that take into account contributions of the self-energy, 
analogous to the scalar case in \eqref{eq:propH}, can also be written down 
for all other massive degrees of freedom in the theory.  

The renormalised fermion propagator is given by
\[
S_R(\slashed{p})\,=\, 
\frac{i}{\slashed{p}-m - \Sigma_{{\rm f} R}(\slashed{p})+i\epsilon}\,,
\label{eq:SRDfin1}
\]
where $m$ is the physical (pole) mass of the fermion and $\Sigma_{{\rm f} R}(\slashed{p})$ is the renormalised
self-energy of the fermion. Details of the definitions of the pole mass, self-energy and the renormalisation constant for the 
fermion field are given in the Appendix. 

It is also straightforward to write down expressions for the resummed propagators for massive vector bosons 
in the Feynman and in the Landau gauge,
 \begin{eqnarray}
{\rm Feynman\,\, gauge}: G_{R}^{\mu\nu} &=& -g^{\mu\nu} \,
\frac{i}{p^2-M^2 -\Sigma_{VR}(p^2)+i \epsilon}
\,,
\label{eq:GRDfinF1}
\\
{\rm Landau\,\, gauge}: \quad  G_{R}^{\mu\nu} &=& -\left( g^{\mu\nu} -\frac{p^\mu p^\nu}{p^2}\right)
\frac{i}{p^2-M^2 -\Sigma_{VR}(p^2)+i \epsilon}
\,.
\label{eq:GRDfinL1}
\end{eqnarray}
In both of these cases the rank-2 tensor on the right hand side of the propagator expressions
does not depend on the mass parameter. The remaining factor is identical to the scalar propagator and we refer the reader to the
Appendix for more detail.

\begin{figure}[t]
\begin{center}
\includegraphics[width=0.4\textwidth]{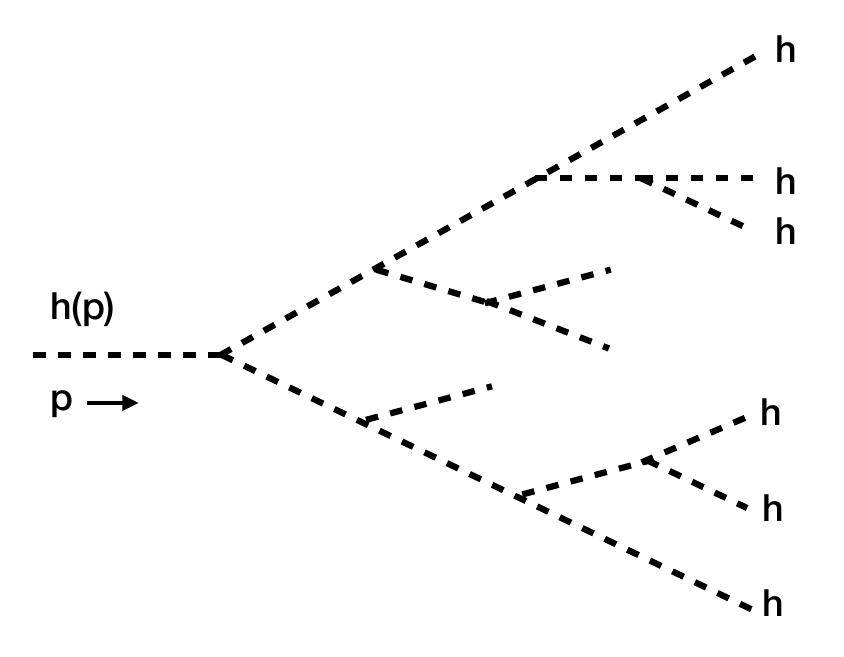}
\hspace{0.5cm}
\includegraphics[width=0.4\textwidth]{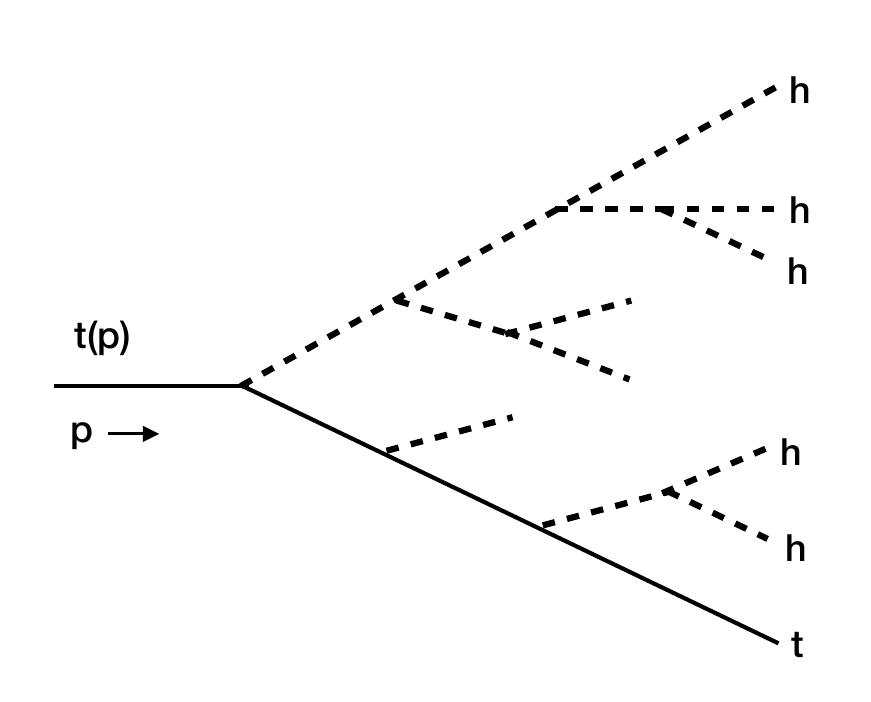}
\end{center}
\caption{Emission of multiple Higgses from the single top and from the single Higgs internal line. The dominant $\sim n!$ 
contribution to the the decay rate at large $n$ comes entirely from the $n\times h$ production which is essentially 
the same for both processes.}
\label{fig:V5}
\end{figure}

All of the propagator expressions above in Eqs.~\eqref{eq:propH} and \eqref{eq:SRDfin1}-\eqref{eq:GRDfinL1}
contain the factor of self-energy in the denominator. As a consequence, at $p^2 \gtrsim E_*^2$ these propagators will 
be suppressed by the decay width $\Gamma(p^2)$ of the propagating state into high-multiplicity final states, as in
Eq.~\eqref{eq:propH2}.
It was explained in Ref.~\cite{Khoze:2017lft}
that if the decay width of a highly virtual off-shell Higgs scalar $h$ into multi-Higgs final states $n\times h$ 
Higgsplodes, it is natural to expect that 
the same Higgsplosion effect will occur for imaginary parts of the self-energies of any degrees of freedom $X$ coupled to the Higgs.
Indeed, at leading order in the loop expansion, there is only a single line difference between the $h \to n\times h$ process and the
 $X\to X+n\times h$, as shown in Fig.~\ref{fig:V5} 
 for the example of $X$ being the top quark. 

\subsection{Loop calculations for observables in the Standard Model}
\label{sec:smloop}
\bigskip

\begin{figure}[t]
\begin{center}
\includegraphics[width=0.4\textwidth]{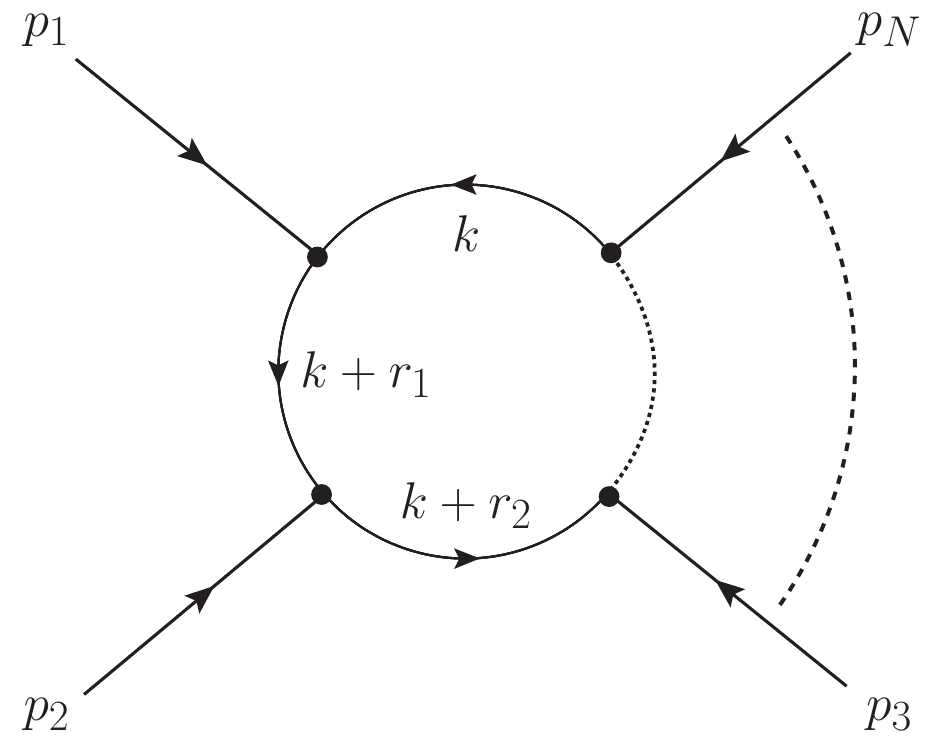}
\end{center}
\caption{General $N$-point function showing the configuration of external and internal momenta flowing through the loop. All momenta are defined to be incoming, and the arrows show the direction of the momenta.}
\label{fig:N-point}
\end{figure}

All the loop-induced processes that we consider arise at the one-loop level. The loop integrals which appear contain various tensor structures and denominators involving external momenta and masses. Following a similar notation to \cite{Denner:2005nn}, the general one-loop $N$-point tensor integral can be written as
\begin{equation}
\label{tensorintegrals}
T_N^{\mu_1\ldots\mu_P}(p_1,\ldots,p_{N-1},m_0,\ldots,m_{N-1}) = \frac{(2\pi\mu)^{4-d}}{i\pi^2}\int d^dk\frac{k^{\mu_1}\ldots k^{\mu_P}}{D_0\ldots D_{N-1}}\,,
\end{equation}
where the denominator contains the propagator factors
\begin{equation}
D_j = (k+r_j)^2-m_j^2+i\varepsilon \qquad , \qquad r_j=\sum_{i=1}^j p_i \qquad , \qquad r_0 = 0\,.
\end{equation}
Fig.~\ref{fig:N-point} shows the general $N$-point function with the configuration of momenta that we use. The $p_i$ are the external momenta of the $N$-point integrals, and the $r_i$ are convenient definitions to simplify the notation. The number of spacetime dimensions in these integrals has been written as $d\equiv 4-\epsilon$, as a precursor to performing dimensional regularisation. The general tensor integrals in Eq.~\eqref{tensorintegrals} can be reduced using the Passarino-Veltman reduction procedure \cite{Passarino:1978jh} to a set of four independent scalar integrals containing no powers of momenta in the numerators. Here, we will focus on observables which reduce to one-point, two-point and three-point scalar integrals. These are:
\begin{align}
\label{A0}
A_0(m_0^2) &= \frac{(2\pi\mu)^{\epsilon}}{i\pi^2}\int d^dk\frac{1}{k^2-m_0^2+i\varepsilon}\,,\\
\label{B0}
B_0(p_1^2,m_0^2,m_1^2) &= \frac{(2\pi\mu)^{\epsilon}}{i\pi^2}\int d^dk\prod_{i=0}^1 \frac{1}{(k+r_i)^2-m_i^2+i\varepsilon}\,,\\
\label{C0}
C_0(p_1^2,p_2^2,r_2^2,m_0^2,m_1^2,m_2^2) &= \frac{(2\pi\mu)^{\epsilon}}{i\pi^2}\int d^dk\prod_{i=0}^2 \frac{1}{(k+r_i)^2-m_i^2+i\varepsilon}\,.
\end{align}
These integrals can be evaluated in dimensional regularisation by Feynman parameterisation, wick rotating into Euclidean space, and then carrying out the $d$-dimensional spherical integrals. The results are:
\begin{align}
A_0(m_0^2) &= m_0^2\left(\frac{2}{\epsilon} - \gamma_\mathrm{E} + \ln4\pi - \ln\frac{m_0^2}{\mu^2} + 1  \right)\,,
\label{eq:A0no}\\
B_0(p_1^2,m_0^2,m_1^2) &=  \frac{2}{\epsilon} - \gamma_\mathrm{E} + \ln4\pi -\int_0^1 dx_1 \ln\frac{M_B^2}{\mu^2}\,,
\label{eq:B0no}\\
C_0(p_1^2,p_2^2,r_2^2,m_0^2,m_1^2,m_2^2) &= -\Gamma(3)\int_0^1dx_1\int_0^{1-x_1}dx_2\frac{1}{2M_C^2}\,.
\label{eq:C0no}
\end{align}
The dummy variables $x_1$ and $x_2$ result from Feynman parameterisation, where we have defined
\begin{equation}
\label{MB}
M_B^2 = (1-x_1)m_1^2+x_1m_2^2+x_1(x_1-1)p_1^2\,,
\end{equation}
for the $B_0$ integral, and
\begin{equation}
\label{MC}
M_C^2 = (1-x_1-x_2)m_1^2+x_1m_2^2+x_2m_3^2+x_1(x_1+x_2-1)p_1^2-x_1x_2p_2^2+x_2(x_1+x_2-1)r_2^2\,,
\end{equation}
for the $C_0$ integral. The $A_0$ tadpole integral and the $B_0$ self-energy integral are UV divergent, which is manifest by the $1/\epsilon$ poles arising from dimensional regularisation. In any physical observable, these UV-divergent contributions must vanish, which is the case for all observables that we consider.


\medskip
\subsection{Procedure for Higgsplosion modifications to loop calculations}
\label{sec:smloop2}
\bigskip


In the Higgsplosion framework, we begin the calculation of the loop integrals for physical processes in the same manner as for the SM. The initial starting point is that we have dimensionally regulated integrals, with the full (infinite) range of momenta allowed. One can then perform Passarino-Veltman decomposition of the tensor integrals to scalar integrals. Higgsplosion is subsequently imposed on the scalar integrals, which formally has the effect of separating the integration into two regions. Below the Higgsplosion scale, we have the unperturbed propagator of the SM and thus there is no change. Above the Higgsplosion scale, the propagator is exponentially suppressed and so the integrand rapidly goes to zero. In practice, this has the effect of restricting the allowed momenta of the integrand to $|k^2|\leq E_*^2$, where $E_*$ is the Higgsplosion scale. The limit in which $E_*\rightarrow \infty$ corresponds to the original dimensionally regulated integrals, which would arise when performing loop calculations in the SM. The integrals can then be Wick rotated into Euclidean space, where the Higgsplosion scale becomes a Euclidean cut-off such that only momenta satisfying $k_{\mathrm{E}}^2 \leq E_*^2$ are allowed. Finally, the integrations can be carried out, with numerical integrations required over the Feynman parameters. Here, we outline the explicit calculations of the UV-divergent integrals $A_0$ and $B_0$, and the UV-finite integral $C_0$.

Starting from Eq.~\eqref{A0}, we perform the Wick rotation to Euclidean momenta and impose Higgsplosion by restricting the range of allowed momenta to $k^2\leq E_*^2$,
\begin{equation}
A_0 \,=\,  -\, \frac{(2\pi\mu)^{\epsilon}}{\pi^2}\int_{k^2\leq E_*^2} d^dk\,\frac{1}{k^2 + m_0^2}\,.
\end{equation}
We will suppress the arguments of the scalar integrals from now on, but the arguments from Eqs.~\eqref{A0}-\eqref{C0} are implicitly assumed.
Carrying out the surface integral over the $d$-dimensional sphere, we get
\begin{equation}
A_0 \,=\, -\, \frac{(2\pi\mu)^\epsilon}{\pi^2}\frac{2\pi^{2-\frac{\epsilon}{2}}}{\Gamma(2-\frac{\epsilon}{2})}\int_0^{E_*}dk\,
\frac{k^{3-\epsilon}}{k^2+m_0^2}\,.
\end{equation}
The restriction on the momenta from Higgsplosion means that this integral is necessarily finite, so dimensional regularisation is no longer required and the $\epsilon\rightarrow 0$ limit can be taken. Finally, carrying out the $k$ integration gives
\begin{equation}
\label{A0final}
A_0 \,=\, m_0^2\log\left(\frac{E_*^2}{m_0^2}+1\right) - E_*^2\,.
\end{equation}
This computation of $A_0$ is identical to our earlier calculation  of the most singular part of the propagator 
in the background of the Brown solution in Section~\ref{sec:proploop}. In fact, $A_0 \,=\, - \, G_I (x,x)$ in Eq.~\eqref{eq:GI}.

The procedure for $B_0$ follows very similar steps. Starting from Eq.~\eqref{B0}, after doing Feynman parameterisation and imposing Higgsplosion, we get
\begin{align}
B_0 &= \frac{(2\pi\mu)^{\epsilon}}{\pi^2}\int_0^1 dx_1\int_{k^2\leq E_*^2} d^dk \frac{1}{(k^2+M_B^2)^2}\,,\nonumber\\
&= \frac{(2\pi\mu)^\epsilon}{\pi^2}\frac{2\pi^{2-\frac{\epsilon}{2}}}{\Gamma(2-\frac{\epsilon}{2})}\int_0^1 dx_1\int_0^{E_*}dk \frac{k^{3-\epsilon}}{(k^2+M_B^2)^2}\,,\nonumber\\
\label{B0final}
&= \int_0^1 dx\left[\log\left(\frac{E_*^2}{M_B^2}+1\right)-\frac{E_*^2}{E_*^2+M_B^2}\right]\,,
\end{align}
with $M_B$ defined in Eq.~\eqref{MB}. Analogously, $C_0$ can be calculated starting from Eq.~\eqref{C0},
\begin{align}
C_0 &= \,-\, \frac{(2\pi\mu)^\epsilon}{\pi^2}\Gamma(3)\int_0^1 dx_1\int_0^{1-x_1} dx_2 \int_{k^2\leq E_*^2} d^dk\frac{1}{(k^2+M_C^2)^3}\,,\nonumber\\
&= \,-\,\frac{(2\pi\mu)^\epsilon}{\pi^2}\Gamma(3)\frac{2\pi^{2-\frac{\epsilon}{2}}}{\Gamma(2-\frac{\epsilon}{2})}\int_0^{E_*}dk\int_0^1 dx_1\int_0^{1-x_1} dx_2 \frac{k^{3-\epsilon}}{(k^2+M_C^2)^3}\,,\nonumber\\
\label{C0final}
&= \,-\,\Gamma(3)\int_0^1 dx_1\int_0^{1-x_1} dx_2\frac{E_*^4}{2M_C^2(E_*^2+M_C^2)^2}\,,
\end{align}
with $M_C$ defined in Eq.~\eqref{MC}. The expressions in Eqs.~\eqref{A0final}-\eqref{C0final} are the final results for the scalar integrals that we use to calculate the effect of Higgsplosion on precision observables. 

We note here that in physical observables, the dependence on the Higgsplosion scale cancels in the $E_*^2 \rightarrow \infty$ limit, which recovers the SM result. This is equivalent to the statement that when using cut-off regularisation, observables are finite in the end. 
The expressions describing the difference in the values of precision observables between the theory with and without Higgsplosion, 
will have the terms $\sim E_*^2$, $\sim \log E_*^2$ and $E_*$-independent terms cancelled. 
Thus, only the terms suppressed by the inverse powers 
of $E_*$ (also allowing additional logarithms) will survive from the integrals in Eqs.~\eqref{A0final}-\eqref{C0final}
in the results for the deviation of the observables from the Standard Model values.
In the language of Section~\ref{sec:loops} this  amounts to the computation of power-suppressed corrections to finite terms $F$ and $B$.


\medskip
\subsection{Results for precision observables}
\label{sec:precisionresults}
\bigskip


Using the expressions from Eqs.~\eqref{A0final}-\eqref{C0final}, we can now compute the effects on precision observables induced by Higgsplosion. All loop-induced observables will receive corrections that scale like $\mathcal{O}(\hat{s}/E_*^2)$. Thus, we focus on observables that can be measured to a high precision. The calculations proceed in accordance with the method outlined in the previous section, where Higgsplosion is imposed after the reduction to scalar integrals. The diagram and amplitude generation is performed by \texttt{FeynArts}~\cite{Hahn:2000kx}, and then the Passarino-Veltman reduction and amplitude squaring is performed by \texttt{FormCalc}~\cite{Hahn:1998yk}. The SM expressions for the loop integrals are evaluated using the \texttt{LoopTools} package.

\subsection{Higgs precision observables}
\label{subsec:ggh}

\begin{figure}[t]
\begin{center}
\includegraphics[width=0.8\textwidth]{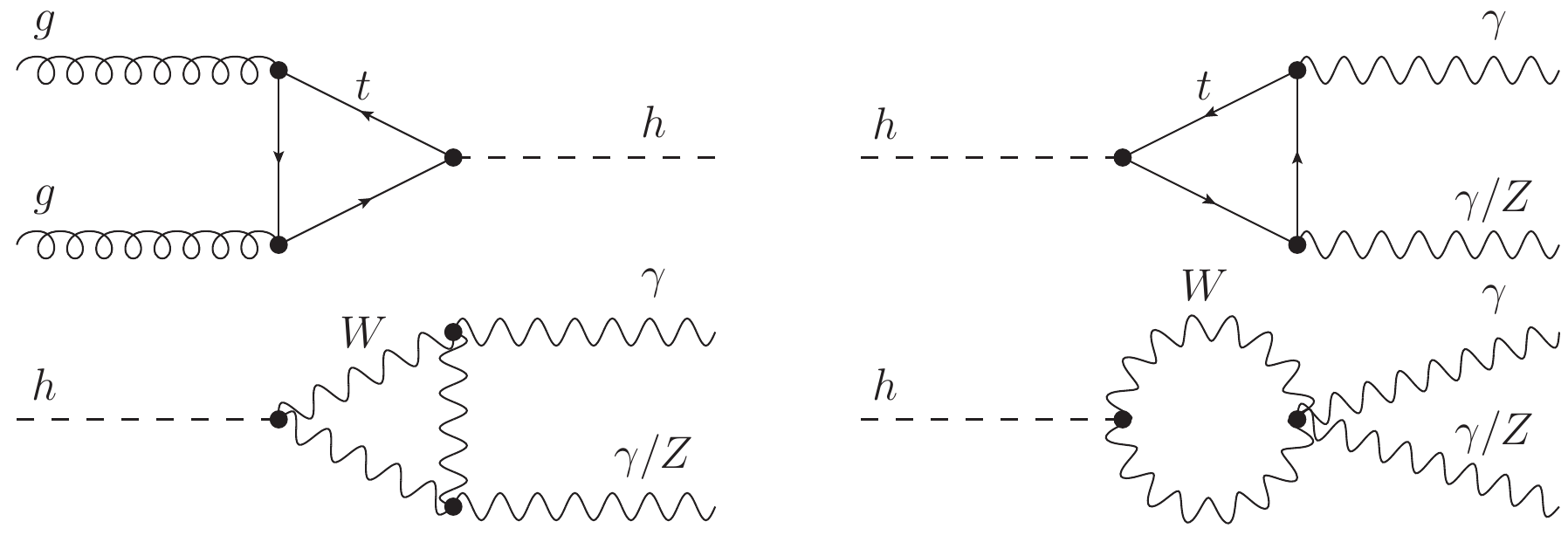}
\end{center}
\caption{Leading-order Feynman diagrams for the Higgs production and decay processes $gg \to h$, $h \to \gamma \gamma$ and $h \to Z \gamma$.}
\label{fig:Higgs-diagrams}
\end{figure}

The first class of observables that we consider are the production and decay of an on-shell Higgs boson via loop-induced leading-order processes, i.e. $gg \to h$, $h \to \gamma \gamma$ and $h \to Z \gamma$. Figure~\ref{fig:Higgs-diagrams} shows the leading-order Feynman diagrams for these processes, capturing the dominant Higgs production mechanism at the LHC, along with two of the cleanest Higgs decay modes. For the production mode, the dominant contribution comes from the top-quark loop, and we neglect the contribution from the lighter fermions. For the decay modes, there are also contributions from $W$-boson loops which interfere destructively with the top-quark loop. 

\begin{figure}[t]
\begin{center}
\includegraphics[width=0.45\textwidth]{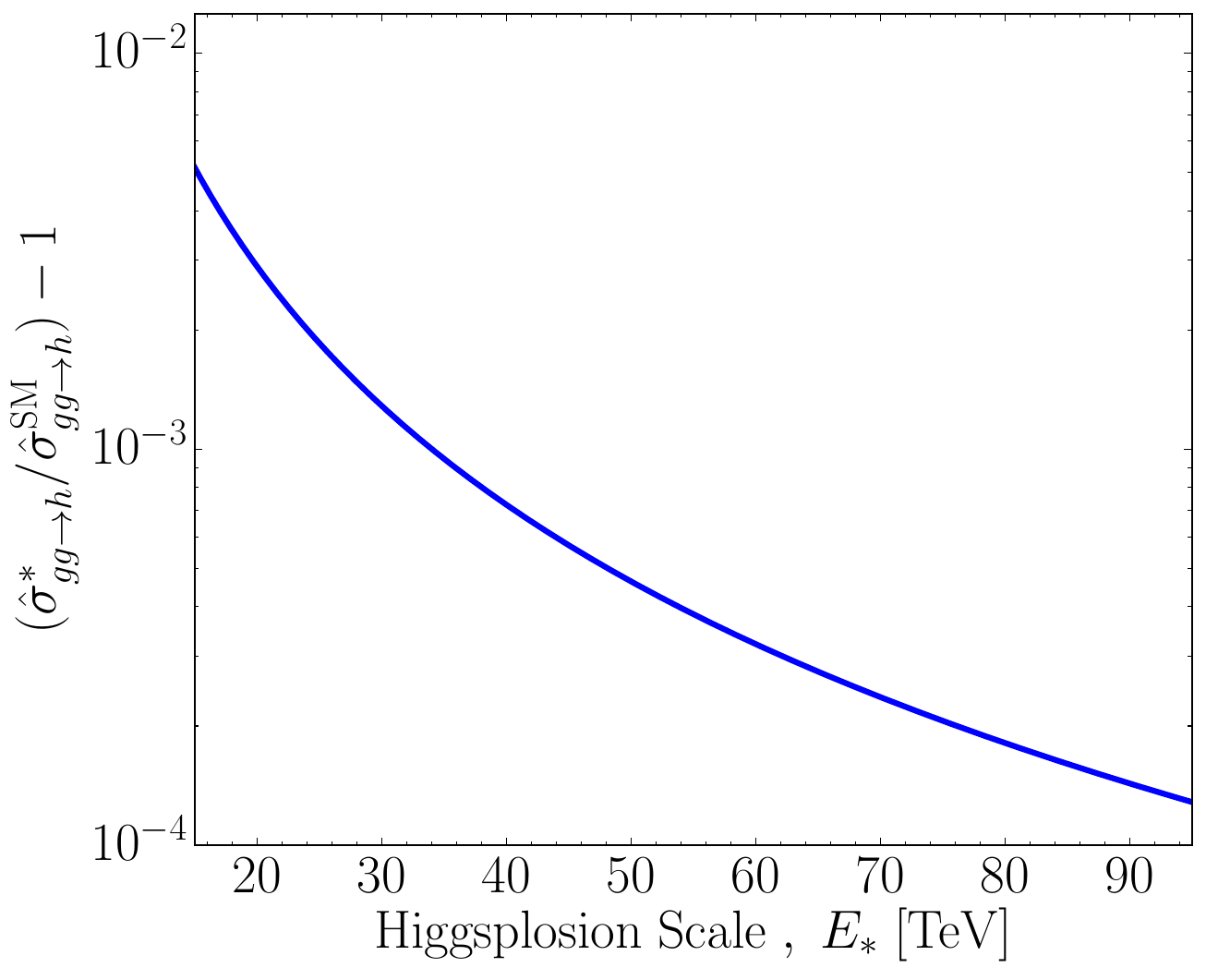}
\includegraphics[width=0.45\textwidth]{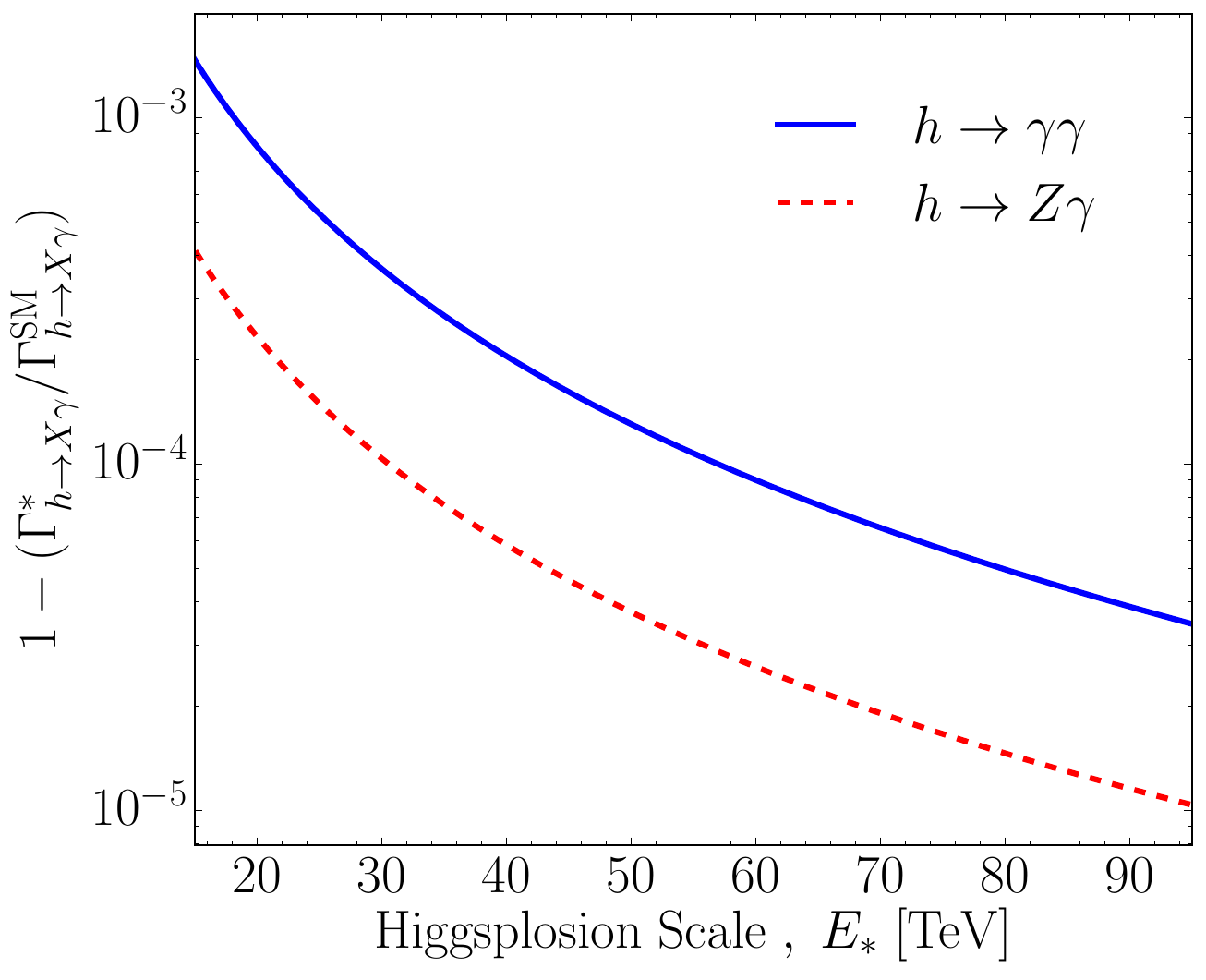}
\end{center}
\caption{Plot of the effect of Higgsplosion on the partonic cross section for $gg\rightarrow h$ (left) and the partial decay width for $h\rightarrow \gamma\gamma$ and $h\rightarrow Z\gamma$ (right) as functions of the Higgsplosion scale $E_*$.}
\label{fig:higgsplots}
\end{figure}

In all three processes the external momentum scale is set by the Higgs mass, $\sqrt{\hat{s}} \simeq m_h$. This is the only external momentum scale that enters into the loop integrals for these processes, which can be decomposed into $C_0$ triangle integrals. To quantify the relative change due to Higgsplosion, we define the quantities $\hat{\sigma}^*$ and $\Gamma^*$  for the partonic cross section and decay width respectively, where the loop integrals have been calculated using the Higgsplosion expressions. These are then compared to their associated SM results, and their relative differences calculated. In Figure~\ref{fig:higgsplots}, we plot the effect of Higgsplosion on the Higgs observables as functions of the Higgsplosion scale, $E_*$.

Current theoretical uncertainties on $gg \to h$ are $\mathcal{O}(10) \%$, irrespective of the center-of-mass energy of the hard process \cite{Contino:2016spe}. Here, a large improvement would be needed to become sensitive to the Higgsplosion scenario. Even if such improvements could be achieved, the predicted experimental sensitivity is $\mathcal{O}(5)\%$ at the LHC with $3000~\mathrm{fb}^{-1}$ \cite{CMS:2013xfa}.
A higher precision can be achieved for the decay $h \to gg$ at a future electron-positron collider. Ref.~\cite{deBlas:2016ojx} gives a predicted precision for $\mathrm{BR}(h \to gg) $ of $1.4\%$ for the FCCee and $3.3\%$ for a 250 GeV ILC.

The expected precision for $\mathrm{BR}(h \to \gamma \gamma)$ is $3.0\%$ at the FCCee. The rate for $\mathrm{BR}(h \to Z \gamma)$ is too small to set tight limits at electron-positron colliders. At the HL-LHC the rate can be limited to $\mathcal{O}(10) \%$ accuracy \cite{No:2016ezr}.
Hence, even assuming strong improvements on the theory uncertainties and data from a future circular electron-positron collider it will not be possible to set limits on the Higgsplosion scale in such measurements of a singly-produced on-shell Higgs boson.

\begin{figure}[t]
\begin{center}
\includegraphics[width=0.45\textwidth]{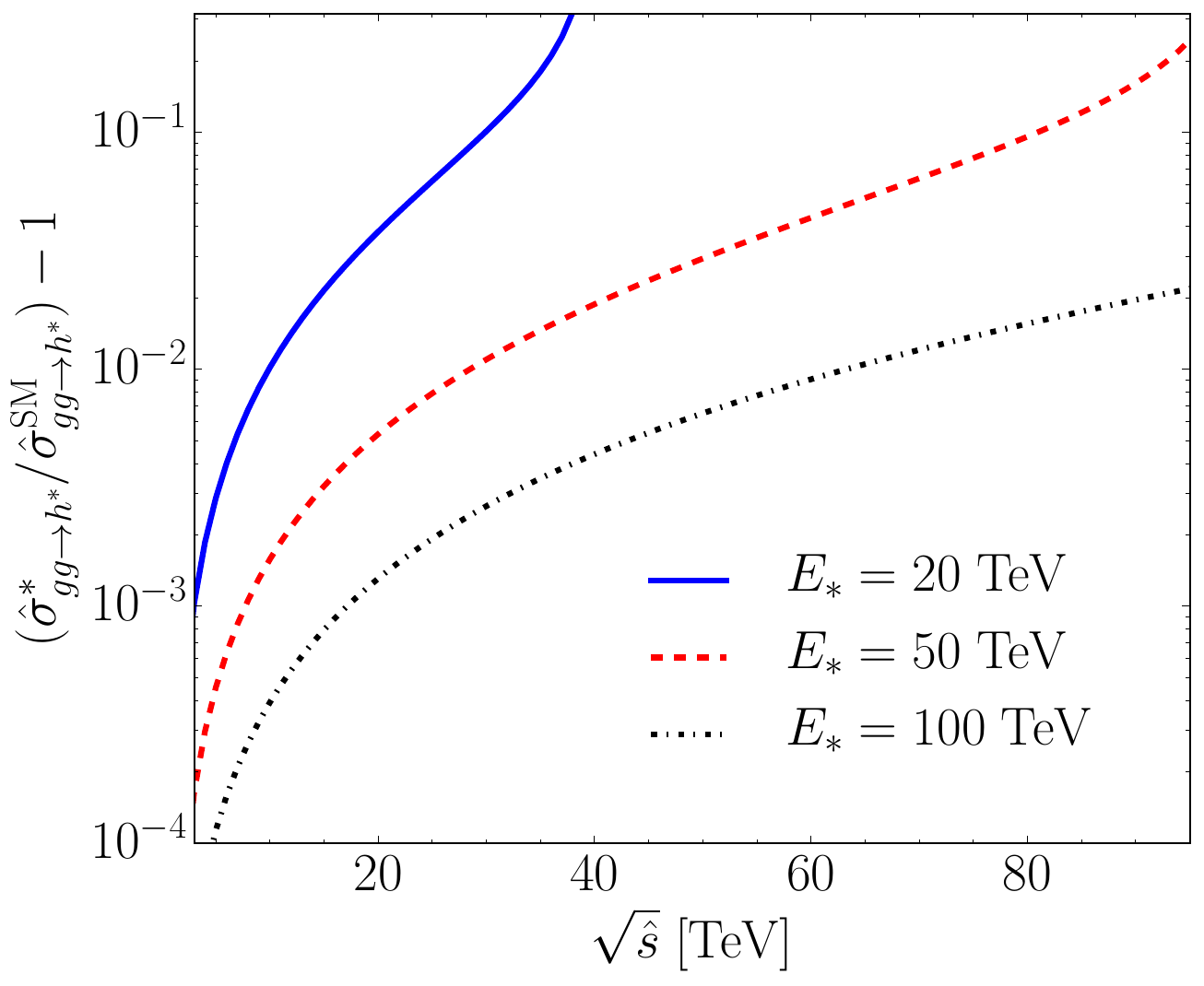}
\end{center}
\caption{Plot of the effect of Higgsplosion on the partonic cross section for $gg \to h^*$ as a function of the centre-of-mass energy $\sqrt{\hat{s}}$. The different curves show different values of the Higgsplosion scale $E_*$.}
\label{fig:gghOffplot}
\end{figure}

A larger effect can be achieved by increasing the interaction scale $\sqrt{\hat{s}}$. In Fig.~\ref{fig:gghOffplot} we show the impact of Higgsplosion on the process $gg\to h^*$, when varying $\sqrt{\hat{s}}$ between $10-90$ TeV. This amounts to producing a Higgs boson far away from its mass-shell or a heavy Higgs boson that could arise in an extension of the Standard Model. The effect becomes $\mathcal{O}(1)$ when $\sqrt{\hat{s}} \simeq 2 E_*$, in close analogy to the $2 m_t$ threshold of the $gg\to h$ process in the Standard Model. The three curves in Fig.~\ref{fig:gghOffplot} correspond to different Higgsplosion scales. As corrections from Higgsplosion scale like $\hat{s}/E_*^2$, the higher the Higgsplosion scale, the large $\sqrt{\hat{s}}$ has to be to achieve an observable effect. This motivates precision studies at a future high-energy collider to test the realisation of Higgsplosion in nature.

\subsection{Flavor observables}
\label{subsec:bsgam}

As Higgsplosion has a direct effect on all loop-induced processes and virtual corrections, flavor observables that have been measured rather precisely could be used to set a limit on the Higgsplosion scale $E_*$. Relevant observables include rare or semileptonic meson decays and Kaon or $B$-meson mixing parameters \cite{Amhis:2016xyh}.

The rate of the rare inclusive decay process $B \to X_s \gamma$ is one of the most important $B$-physics observables as it sets stringent constraints on the parameter space of various extensions of the SM \cite{Buras:1993xp}. At lowest order it can be described by the transition $b \to s \gamma$. The effective Hamiltonian for this decay is usually expressed as \cite{Ali:1990vp} 
\begin{equation}
\mathcal{H}_\mathrm{eff} = -\frac{4G_F}{\sqrt{2}}V_{tb}V^*_{ts}\sum_{i=1}^8 C_i(\mu) \mathcal{O}_i(\mu)\,,
\end{equation}
where $V_{ij}$ are elements of the CKM matrix, $G_F$ is the Fermi constant and $\mu$ is the scale at which the Wilson coefficients $C_i(\mu)$ are evaluated at.

\begin{figure}[t]
\begin{center}
\includegraphics[width=0.45\textwidth]{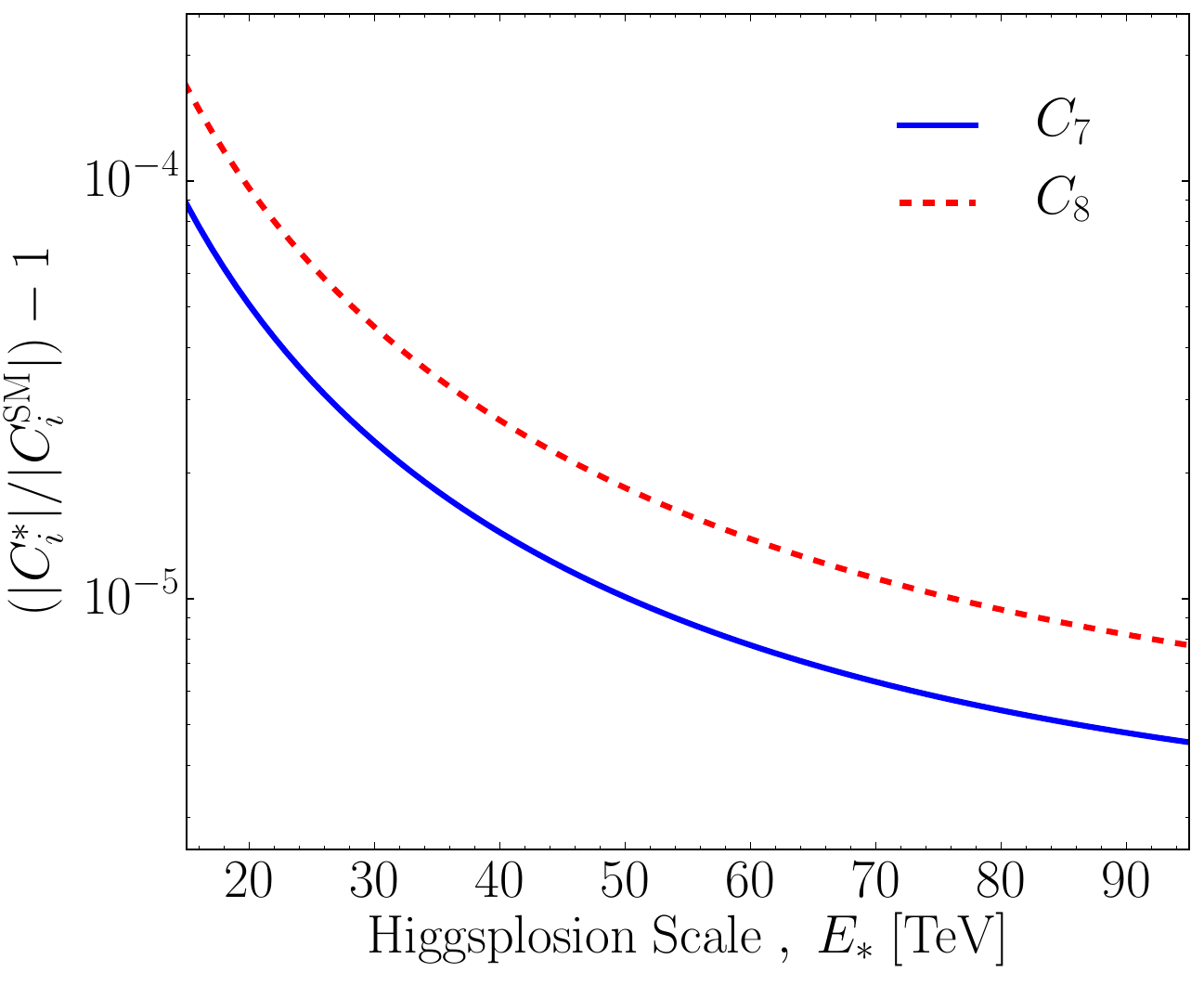}
\end{center}
\caption{Plot of the effect of Higgsplosion on the Wilson coefficients $C_7$ and $C_8$ for $B \to X_s \gamma$ as functions of the Higgsplosion scale $E_*$.}
\label{fig:flavplots}
\end{figure}

The effect of Higgsplosion is predominantly encoded in the Wilson coefficients. Their relative change from the SM directly modifies the decay rate of $B \to X_s \gamma$ by the same amount. Here, we will focus on the coefficients $C_7$ and $C_8$, which are associated with the operators
\begin{align}
\mathcal{O}_7 &= \frac{e}{16\pi^2}m_b \braket{s_j|\omega_+\sigma_{\mu\nu}|b_i}\delta_{ij}F^{\mu\nu}\,,\\
\mathcal{O}_8 &= \frac{g_s}{16\pi^2}m_b \braket{s_j|\omega_+\sigma_{\mu\nu}|b_i}T^a_{ij}G^{\mu\nu}_a\,.
\end{align}
They are calculated from the partonic transitions $b \to s \gamma$ and $b \to sg$ respectively. Fig.~\ref{fig:flavplots} shows the Wilson coefficients $C_7$ and $C_8$ as functions of the Higgsplosion scale, $E_*$. 

We find that the Higgsplosion modifications of $B \to X_s \gamma$ compared to the SM are small and are, even for a low Higgsplosion scale of $E_* \simeq 30$ TeV, unobservable, taking theoretical and experimental uncertainties into account. Current measurements of $\mathrm{BR}(B \to X_s \gamma) \simeq (335 \pm 15) \cdot 10^{-6}$ \cite{Chen:2001fja,Limosani:2009qg,Lees:2012ym} are at the level of $5 \%$ accuracy only.

\subsection{Anomalous magnetic dipole moment of the electron and muon}
\label{subsec:gmin2}

The anomalous magnetic dipole moment of the electron is one of the great successes of twentieth century physics, and QED specifically. The precision with which theoretical predictions agree with experimental measurement is about one part in a trillion, which is unprecedented for many areas of physics. However, for a muon there is a discrepancy with the SM prediction \cite{Bennett:2006fi, Jegerlehner:2009ry, Hagiwara:2011af,Davier:2010nc}

\begin{equation}
\Delta a_\mu = a^{\mathrm{exp}}_\mu - a^{\mathrm{theory}}_\mu \simeq 2.90 \cdot 10^{-9}\,,
\end{equation}
which may be a sign of new physics.

\begin{figure}[t]
\begin{center}
\includegraphics[width=0.45\textwidth]{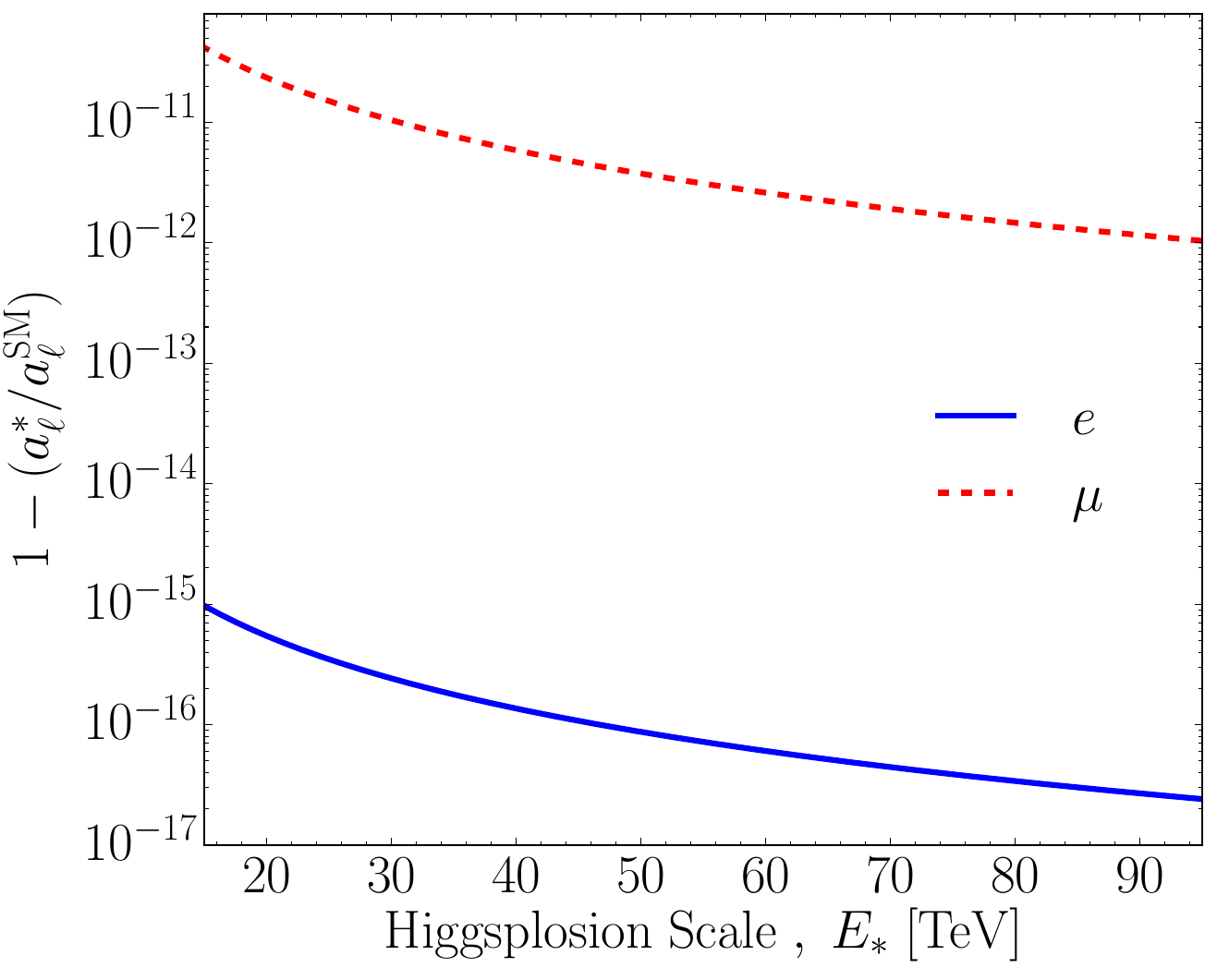}
\end{center}
\caption{Plot of the effect of Higgsplosion on the anomalous magnetic moment of the electron and muon as functions of the Higgsplosion scale $E_*$.}
\label{fig:gminus2plot}
\end{figure}

The anomalous magnetic moment of a fermion quantifies how much its magnetic moment $g$ differs from its classical value, which is predicted by the Dirac equation. This is quantified by the expression $a = (g-2)/2$, where $a$ is the anomalous magnetic moment. In perturbative QED, the tree level result corresponds to the vertex interaction of a charged lepton and photon at zero momentum transfer, and recovers the classical prediction. The radiative corrections to this vertex can in general be described by the form-factors $F_1$ and $F_2$,
\begin{equation}
\Gamma^{\mu} = F_1(q^2)\gamma^\mu + F_2(q^2)\frac{i\sigma^{\mu\nu}q_\nu}{2m}\,.
\end{equation}
The anomalous magnetic moment is then given by $F_2(0)$. We calculate the one-loop contributions to the anomalous magnetic moment of the electron and muon and their deviations due to Higgsplosion. The one-loop result for a charged lepton $\ell$ can be compactly written as
\begin{equation}
a_\ell = \frac{\alpha}{4\pi}\left[2B_0(m_\ell^2,0,m_\ell^2)-B_0(0,m_\ell^2,m_\ell^2)-B_0(0,0,m_\ell^2)-1\right]\,.
\end{equation}
In the SM at one loop, the mass dependence in the $B_0$ integrals cancels so the anomalous magnetic moment is $a_\ell = \alpha / (2\pi)$ for all charged leptons. However, in Higgsplosion the mass dependence remains and changes are induced via the $B_0$ integrals. The effect of Higgsplosion is shown by the plot in Fig.~\ref{fig:gminus2plot}. We find that the sensitivity on $a_\mu$ has to be improved by at least two orders of magnitude to be able to set a meaningful limit on the Higgsplosion scale $E_*$.

The anomalous magnetic moment of the electron has been measured to be \cite{Hanneke:2010au}
\begin{equation}
a_e^{\mathrm{exp}} = 11596521807.3(2.8) \cdot 10^{-13}\,,
\end{equation}
with an experimental uncertainty of $\delta a_e^{\mathrm{exp}} = 2.8 \cdot 10^{-13}$. Such high precision allows one to set limits on a wide range of new physics models \cite{Giudice:2012ms}. However, as the relative changes to the anomalous magnetic moments of the electron and muon induced by Higgsplosion are related by
\begin{equation}
\frac{1-a_e^*/a_e^{\mathrm{SM}}}{1-a_\mu^*/a_\mu^{\mathrm{SM}}} \approx \frac{m^2_e}{m_\mu^2}\,,
\end{equation}
the increased precision for the electron compared with the muon does not translate into a better limit on the Higgsplosion scale.


\medskip
\section{Discussion and Conclusions}
\label{sec:concl}

The motivation and main task of this paper was to provide the first detailed study of how
the characteristic energy scale of Higgsplosion affects the loop integrals in Higgsploding theories.

There are dramatic consequences of Higgsplosion for the theory dynamics at energy scales in the UV. At the same
time, we have shown that at currently accessible energies the effects of the Higgsplosion scale remain very small. However, $\mathcal{O}(1)$ effects can be achieved for loop-induced processes or virtual corrections if the interactions are probed close to $\sim 2 E_*$. This effect can be immediately seen in Fig.~\ref{fig:gghOffplot} for $gg \to h^*$, but it has much broader applicability. It applies to all Standard Model processes, such as Drell-Yan, multi-jet production or the production of electroweak gauge bosons.

\medskip

\noindent The main features of Higgsplosion that have been explored in \cite{Khoze:2017tjt, Khoze:2017lft} and the present paper are:
\begin{enumerate}
\item There are no UV divergences in the theory.
\item The running of the coupling and mass parameters of the theory flattens out when the RG scale $\mu$ exceeds the Higgsplosion scale $E_*$ and the parameters reach their UV fixed points. There are no Landau poles and no couplings become negative. Therefore, the Higgsploding theory is asymptotically safe.
\item There are no hierarchically large contributions to the Higgs boson mass arising from potentially very heavy  degrees of freedom
with masses above $E_*$. 
\item More generally, all particle states at energies or virtualities above $E_*$ rapidly decay into multiple relatively soft Higgs
bosons and massive vector bosons. Essentially, at $\sqrt{s}>E_*$, all elementary particle states including the Higgs itself become 
composite states involving high multiplicities $n\sim \sqrt{s}/m_h$ of soft electroweak quanta.
\item The self-consistency of the computational formalism, in particular the applications of the
semi-classical formalism of \cite{Son:1995wz} to Higgsplosion \cite{Khoze:2017tjt,Khoze:2017ifq}, requires that the exponentiation of the loop corrections to the
tree-level Higgsploding amplitudes, originally proved in \cite{Libanov:1994ug}, remains unaffected when the loop integrations are 
modified by the presence of the Higgsplosion scale $E_*$.
In Section~\ref{sec:loops} we demonstrated that this is indeed the case.
\item In Section~\ref{sec:qft} we followed this up with the loop formalism in the presence of Higgsplosion and
computed a set of Standard Model precision observables. We found that the effects of the Higgsplosion scale on these observables 
 are small and remain unobservable at currently available energies.
\item This of course does not affect the possibility of direct observation of Higgsplosion at energies $\sqrt{s}$ above
the $E_*$ threshold \cite{Khoze:2014kka,Degrande:2016oan},
which would potentially be achievable at future hadron colliders in the 100 TeV range, 
or in the early Universe (e.g.~the decays and interactions of the inflaton during reheating).
\end{enumerate}

It was pointed out in  \cite{Khoze:2017tjt} that the direct Higgsplosive 
production of multiple Higgs bosons in very
high energy collisions with $\sqrt{s}>E_*$ does not result in the breakdown of perturbative unitarity even when 
the rates appear to grow exponentially with energy.
The computation of the cross section of physical processes, such as gluon fusion  $gg \to n\times h$ going through an intermediate virtual Higgs boson(s) produced in the $s$-channel,
$gg \to h^*\to n\times h$,
requires the use of the dressed propagators for the intermediate $h^*$.
This results in Higgspersion, i.e. a well-behaved cross section for arbitrary $n$ up to very high energies~\cite{Khoze:2017tjt},
\[ \sigma^{\Delta}_{gg \to n\times h} \, \sim \, 
y_t^{2}   
m_t^2  \log^4\left(\frac{m_t}{\sqrt{p^2}}\right) \,\times\, \frac{1}{p^4+m_h^4{\cal R}^2}
\,\times\, 
{\cal R}_n\,,
\label{eq:higgsper}
\]
and thus
\[
\sigma_{gg \to n\times h}\, \sim \, 
\begin{cases}
\,\,{\cal R} & :\,\,{\rm for}\,\, \sqrt{s}\le E_*\,\,{\rm where} \,\,  {\cal R} \lesssim 1\\
\,\,1/{\cal R}\to 0 & :\,\,{\rm for}\,\,  \sqrt{s}\ge E_*\,\,{\rm where} \,\, {\cal R} \gg 1
\end{cases}\, .
\label{eq:poly_ev_odd}
\]
Hence, by avoiding a breakdown of perturbative unitarity in multi-boson production, the theory can retain consistency and predictivity to much higher, technically even unlimited, energy scales.

It is important to keep in mind that the expression for the cross section in \eqref{eq:poly_ev_odd} does not imply
that the physical cross section once again becomes exponentially small when $p^2$ or $s$ is $\gg E_*^2$.
This expression was derived under the assumption that all the energy of the collision goes into producing 
as many soft quanta as kinematically possible. However, at energies exceeding $E_*$, this is no longer the case.
Instead it is more advantageous for the initial highly energetic particles to emit one or few hard quanta
to lower the energy from $\sqrt{s}$ down to $\sim E_*$. The scattering then proceeds by emitting mostly soft quanta.
Thus the correct behaviour for the Higgsplosion cross section is that it saturates at high energies,
\[
\sigma_{gg \to n\times h}\, \sim \, 
\begin{cases}
\,\,{\cal R} & :\,\,{\rm for}\,\, \sqrt{s}\ll E_*\,\,{\rm where} \,\, {\cal R} \ll 1\\
\,\,1 & :\,\,{\rm for}\,\, \sqrt{s}\ge E_*\,\,{\rm where} \,\, {\cal R} \gg 1
\end{cases}\, .
\label{eq:poly_ev_new}
\]
This leaves the possibility of direct Higgsplosion processes being observable at energies above the Higgsplosion scale.


\bigskip

\acknowledgments
We would like to thank Claude Duhr for valuable discussions. 

 \bigskip
 \startappendix
\Appendix{Dyson-resummed propagators for fermions and massive vector bosons}
\label{App:propFV}

The purpose of the discussion that follows is to show how the Dyson resummation of the scalar field propagator  
\eqref{eq:propH}, outlined in section~\ref{sec:Dyson},
applies to massive fermions and vector bosons
and to derive the expressions in Eqs.~\eqref{eq:SRDfin1}-\eqref{eq:GRDfinL1}
for the propagators of massive fermions and vector bosons quoted in section~\ref{sec:propFV}.

The unrenormalised fermion propagator with the bare mass $m_0$,
\[
S(\slashed{p})\,=\, 
\frac{i}{\slashed{p}-m_0 -\Sigma_{\rm f}(\slashed{p}) + i \epsilon}
\,,
\label{eq:drexpFS}
\]
can be understood  in perturbation theory as an 
infinite expansion in terms of bare propagators with the insertions of the self-energy, 
\[
\frac{i}{\slashed{p}-m_0 -\Sigma_{\rm f}(\slashed{p})} \,=\, 
\frac{i}{\slashed{p}-m_0} \,+\, 
\frac{i}{\slashed{p}-m_0}\, \sum_{n=1}^\infty
 \left(-i \Sigma_{\rm f}(\slashed{p})\, \frac{i}{\slashed{p}-m_0}\right)^n
\,,
\label{eq:drexpF}
\]
in complete analogy to the scalar propagator formula \eqref{eq:dres}. 
 The fermion self-energy is denoted $\Sigma_{\rm f}(\slashed{p})$, the momentum variable $\slashed{p}$
is the usual contraction with the gamma matrices, $p_\mu \gamma^\mu$, and it is also understood that 
$\slashed{p}^2= p_\mu p^\mu = p^2$.

The physical (or pole) mass $m$ of the fermion is defined in analogy to the scalar case in \eqref{eq:mdef},
as the value of the momentum variable $\slashed{p}=m \cdot 1 = m$ for which the denominator in \eqref{eq:drexpFS} is zero,
\[
m\,=\, m_0 \, +\,  \Sigma_{\rm f}(m)\,.
\label{eq:mdefF}
\]
The expressions for the renormalised fermion propagator and the renormalised self-energy
 ({\it cf.}~\eqref{eq:RpropS}-\eqref{eq:Sigmasub}) are, see for example \cite{Itzykson:1980rh},
 \begin{eqnarray}
S_R(\slashed{p}) &=& Z_2^{\,\,-1} \, S(\slashed{p})
\, , 
\label{eq:RpropF} \\
\Sigma_{{\rm f} R}(\slashed{p}) &=& Z_2 \, \left(\Sigma_{\rm f}(\slashed{p})-\Sigma_{\rm f}(m)\right)\,-\, 
(Z_2-1)(\slashed{p}-m)\,,
\label{eq:SigmasubF}
\end{eqnarray}
where $Z_2$
is the renormalisation constant of the fermion field.
The right-hand side of \eqref{eq:SigmasubF} is nothing more than a $Z_2 \Sigma_{\rm f}(\slashed{p})$ renormalised
combination with the appropriate subtractions of factors of $p$ and $m$, in analogy to the scalar self-energy renormalisation
condition \eqref{eq:Sigmasub}. What makes the definition \eqref{eq:SigmasubF} particularly useful is the fact that it can 
be rearranged to give
\[
\slashed{p}-m - \Sigma_{{\rm f} R}(\slashed{p})\,=\, Z_2 \, \left(\slashed{p}-m_0 - \Sigma_{{\rm f} }(\slashed{p})\right)\,,
\label{eq:propFRuse}
\] 
where we have also used the pole mass equation \eqref{eq:mdefF}.

The relation \eqref{eq:propFRuse} results in the following simple expression for the renormalised 
Dyson resummed fermion propagator,
\[
S_R(\slashed{p})\,=\, 
\frac{i}{\slashed{p}-m - \Sigma_{{\rm f} R}(\slashed{p})+i\epsilon}\,. 
\label{eq:SRDfin}
\]
The definition of the pole mass implies that the above expression for the renormalised propagator must have 
a pole at $\slashed{p}=m$, and the residue must be equal to $i$. These two conditions give,
\begin{eqnarray}
\Sigma_{{\rm f} R} (m) &=& 0\,, \label{eq:1stF}  \\
\left(d \Sigma_{{\rm f}R}/d\slashed{p}\right)|_{\slashed{p}=m} &=& 0\,. \label{eq:2ndF}
\end{eqnarray}
We now check what these conditions imply for the defining expression for $\Sigma_{{\rm f} R}(\slashed{p})$ in 
\eqref{eq:SigmasubF}. To do this, we Taylor expand both sides of \eqref{eq:SigmasubF} around $\slashed{p}=m$.
By evaluating \eqref{eq:SigmasubF} at $\slashed{p}=m$, we find that $\Sigma_{{\rm f} R} (m) =0$, which is precisely 
the first condition in Eq.~\eqref{eq:1stF}. The first derivative of \eqref{eq:SigmasubF} gives,
\[
\left(d \Sigma_{{\rm f}R}/d\slashed{p}\right)|_{\slashed{p}=m} \,=\, 
Z_2 \left(d \Sigma_{{\rm f}}/d\slashed{p}\right)|_{\slashed{p}=m} +1 -Z_2\,,
\]
and imposing the second condition, \eqref{eq:2ndF}, we set the right-hand side to zero, which results in,
\[
Z_2 \,=\, \left(1 -\left.\frac{d \Sigma_{\rm f}}{d\slashed{p}}\right\vert_{\slashed{p}=m}\right)^{-1}\,.
\label{eq:mass1F}
\]
This expression fixes the renormalisation constant $Z_2$ in terms of the slope of the fermionic self-energy at 
$\slashed{p}=m$. This is analogous to the scalar field renormalisation constant definition we discussed before in 
\eqref{eq:mass1}.	

\medskip

Having derived the expressions for full quantum propagators of massive scalars  \eqref{eq:propH} and 
massive fermions \eqref{eq:SRDfin}, we can follow the same route to establish the Dyson-resummed 
propagators of massive vector bosons. Our starting point is the expression for the bare propagator of a vector boson with a bare mass $M_0$, 
\[
G_0^{\mu\nu}\,=\, -\left( g^{\mu\nu} +(\xi-1)\frac{p^\mu p^\nu}{p^2-\xi M_0^2}\right)
\frac{i}{p^2-M_0^2 +i \epsilon}
\,,
\label{eq:GRDfin}
\]
where  $\xi$ is the gauge-fixing parameter. The choice $\xi=1$ gives the Feynman gauge and $\xi=0$ is the Landau gauge.
In these two cases, the rank-2 tensor appearing on the right-hand side of the bare propagator is particularly simple as it
does not depend on the mass parameter. This allows us to readily resum a geometric progression series involving a
sequence of bare propagators with insertions of the vector boson self-energy. 

The renormalised Dyson propagators in the Feynman and the Landau gauge are thus given by,
 \begin{eqnarray}
{\rm Feynman\,\, gauge}: \quad G_{R}^{\mu\nu} &=& -g^{\mu\nu} \,
\frac{i}{p^2-M^2 -\Sigma_{VR}(p^2)+i \epsilon}
\,,
\label{eq:GRDfinF}\\
{\rm Landau\,\, gauge}: \quad  G_{R}^{\mu\nu}&=& -\left( g^{\mu\nu} -\frac{p^\mu p^\nu}{p^2}\right)
\frac{i}{p^2-M^2 -\Sigma_{VR}(p^2)+i \epsilon}
\,,
\label{eq:GRDfinL}
\end{eqnarray}
where $M$ is the pole mass and $\Sigma_{VR} (p^2)$ is the renormalised self-energy of the vector boson.

The pole mass $M$ is related to the bare mass $M_0$ in exactly the same way 
as in the scalar case~\eqref{eq:mdef},
\[
M^2 - M_0^2 -  \Sigma_V(M^2)\,=\, 0 \,.
\label{eq:mdefV}
\]
The renormalised self-energy is defined via,
\[
\Sigma_{VR} (p^2)\,=\, Z_1 \,\left(\Sigma_V(p^2)-\Sigma(M^2)- \Sigma'_V (M^2) (p^2-M^2)\right)\,,
\label{eq:Sigmasub0}
\]
where $\Sigma_V(p^2)$ is the unrenormalised self-energy,
and $Z_1$ is the field-strength renormalisation constant of the vector field. It is given by,
\[
Z_1\,=\, \left(1 -\left.\frac{d \Sigma_V}{dp^2}\right\vert_{p^2=M^2}\right)^{-1}\,.
\label{eq:mass1V}
\]
The expressions in \eqref{eq:Sigmasub0}, \eqref{eq:mass1V} are
identical to those in Eqs.~\eqref{eq:Sigmasub} and \eqref{eq:mass1}
in the scalar case.

\bibliographystyle{JHEP}
\bibliography{references}

\providecommand{\href}[2]{#2}\begingroup\raggedright\begin{thebibliography}{10}

\bibitem{Khoze:2017tjt}
V.~V. Khoze and M.~Spannowsky, \emph{{Higgsplosion: Solving the Hierarchy
  Problem via rapid decays of heavy states into multiple Higgs bosons}},
  \href{https://doi.org/10.1016/j.nuclphysb.2017.11.002}{\emph{Nucl. Phys.}
  {\bfseries B926} (2018) 95--111},
  [\href{https://arxiv.org/abs/1704.03447}{{\ttfamily 1704.03447}}].

\bibitem{Khoze:2017lft}
V.~V. Khoze and M.~Spannowsky, \emph{{Higgsploding universe}},
  \href{https://doi.org/10.1103/PhysRevD.96.075042}{\emph{Phys. Rev.}
  {\bfseries D96} (2017) 075042},
  [\href{https://arxiv.org/abs/1707.01531}{{\ttfamily 1707.01531}}].

\bibitem{Manton:1983nd}
N.~S. Manton, \emph{{Topology in the Weinberg-Salam Theory}},
  \href{https://doi.org/10.1103/PhysRevD.28.2019}{\emph{Phys. Rev.} {\bfseries
  D28} (1983) 2019}.

\bibitem{Klinkhamer:1984di}
F.~R. Klinkhamer and N.~S. Manton, \emph{{A Saddle Point Solution in the
  Weinberg-Salam Theory}},
  \href{https://doi.org/10.1103/PhysRevD.30.2212}{\emph{Phys. Rev.} {\bfseries
  D30} (1984) 2212}.

\bibitem{Brown:1992ay}
L.~S. Brown, \emph{{Summing tree graphs at threshold}},
  \href{https://doi.org/10.1103/PhysRevD.46.R4125}{\emph{Phys. Rev.} {\bfseries
  D46} (1992) R4125--R4127},
  [\href{https://arxiv.org/abs/hep-ph/9209203}{{\ttfamily hep-ph/9209203}}].

\bibitem{Argyres:1992np}
E.~N. Argyres, R.~H.~P. Kleiss and C.~G. Papadopoulos, \emph{{Amplitude
  estimates for multi - Higgs production at high-energies}},
  \href{https://doi.org/10.1016/0550-3213(93)90140-K}{\emph{Nucl. Phys.}
  {\bfseries B391} (1993) 42--56}.

\bibitem{Voloshin:1992rr}
M.~B. Voloshin, \emph{{Estimate of the onset of nonperturbative particle
  production at high-energy in a scalar theory}},
  \href{https://doi.org/10.1016/0370-2693(92)90901-F}{\emph{Phys. Lett.}
  {\bfseries B293} (1992) 389--394}.

\bibitem{Libanov:1994ug}
M.~V. Libanov, V.~A. Rubakov, D.~T. Son and S.~V. Troitsky,
  \emph{{Exponentiation of multiparticle amplitudes in scalar theories}},
  \href{https://doi.org/10.1103/PhysRevD.50.7553}{\emph{Phys. Rev.} {\bfseries
  D50} (1994) 7553--7569},
  [\href{https://arxiv.org/abs/hep-ph/9407381}{{\ttfamily hep-ph/9407381}}].

\bibitem{Voloshin:1992nu}
M.~B. Voloshin, \emph{{Summing one loop graphs at multiparticle threshold}},
  \href{https://doi.org/10.1103/PhysRevD.47.R357}{\emph{Phys. Rev.} {\bfseries
  D47} (1993) R357--R361},
  [\href{https://arxiv.org/abs/hep-ph/9209240}{{\ttfamily hep-ph/9209240}}].

\bibitem{Smith:1992rq}
B.~H. Smith, \emph{{Summing one loop graphs in a theory with broken symmetry}},
  \href{https://doi.org/10.1103/PhysRevD.47.3518}{\emph{Phys. Rev.} {\bfseries
  D47} (1993) 3518--3520},
  [\href{https://arxiv.org/abs/hep-ph/9209287}{{\ttfamily hep-ph/9209287}}].

\bibitem{Voloshin:2017flq}
M.~B. Voloshin, \emph{{Loops with heavy particles in production amplitudes for
  multiple Higgs bosons}},
  \href{https://doi.org/10.1103/PhysRevD.95.113003}{\emph{Phys. Rev.}
  {\bfseries D95} (2017) 113003},
  [\href{https://arxiv.org/abs/1704.07320}{{\ttfamily 1704.07320}}].

\bibitem{Son:1995wz}
D.~T. Son, \emph{{Semiclassical approach for multiparticle production in scalar
  theories}}, \href{https://doi.org/10.1016/0550-3213(96)00386-0}{\emph{Nucl.
  Phys.} {\bfseries B477} (1996) 378--406},
  [\href{https://arxiv.org/abs/hep-ph/9505338}{{\ttfamily hep-ph/9505338}}].

\bibitem{Libanov:1997nt}
M.~V. Libanov, V.~A. Rubakov and S.~V. Troitsky, \emph{{Multiparticle processes
  and semiclassical analysis in bosonic field theories}},
  \href{https://doi.org/10.1134/1.953038}{\emph{Phys. Part. Nucl.} {\bfseries
  28} (1997) 217--240}.

\bibitem{Gorsky:1993ix}
A.~S. Gorsky and M.~B. Voloshin, \emph{{Nonperturbative production of
  multiboson states and quantum bubbles}},
  \href{https://doi.org/10.1103/PhysRevD.48.3843}{\emph{Phys. Rev.} {\bfseries
  D48} (1993) 3843--3851},
  [\href{https://arxiv.org/abs/hep-ph/9305219}{{\ttfamily hep-ph/9305219}}].

\bibitem{Khoze:2017ifq}
V.~V. Khoze, \emph{{Multiparticle production in the large $\lambda n$ limit:
  realising Higgsplosion in a scalar QFT}},
  \href{https://doi.org/10.1007/JHEP06(2017)148}{\emph{JHEP} {\bfseries 06}
  (2017) 148}, [\href{https://arxiv.org/abs/1705.04365}{{\ttfamily
  1705.04365}}].

\bibitem{Dyson:1949ha}
F.~J. Dyson, \emph{{The S matrix in quantum electrodynamics}},
  \href{https://doi.org/10.1103/PhysRev.75.1736}{\emph{Phys. Rev.} {\bfseries
  75} (1949) 1736--1755}.

\bibitem{Schwinger:1951ex}
J.~S. Schwinger, \emph{{On the Green's functions of quantized fields. 1.}},
  \href{https://doi.org/10.1073/pnas.37.7.452}{\emph{Proc. Nat. Acad. Sci.}
  {\bfseries 37} (1951) 452--455}.

\bibitem{Schwinger:1951hq}
J.~S. Schwinger, \emph{{On the Green's functions of quantized fields. 2.}},
  \href{https://doi.org/10.1073/pnas.37.7.455}{\emph{Proc. Nat. Acad. Sci.}
  {\bfseries 37} (1951) 455--459}.

\bibitem{Gainer:2017jkp}
J.~S. Gainer, \emph{{Measuring the Higgsplosion Yield: Counting Large Higgs
  Multiplicities at Colliders}},
  \href{https://arxiv.org/abs/1705.00737}{{\ttfamily 1705.00737}}.

\bibitem{Khoze:2014kka}
V.~V. Khoze, \emph{{Perturbative growth of high-multiplicity W, Z and Higgs
  production processes at high energies}},
  \href{https://doi.org/10.1007/JHEP03(2015)038}{\emph{JHEP} {\bfseries 03}
  (2015) 038}, [\href{https://arxiv.org/abs/1411.2925}{{\ttfamily 1411.2925}}].

\bibitem{Rubakov:1991fb}
V.~A. Rubakov and P.~G. Tinyakov, \emph{{Towards the semiclassical
  calculability of high-energy instanton cross-sections}},
  \href{https://doi.org/10.1016/0370-2693(92)91859-8}{\emph{Phys. Lett.}
  {\bfseries B279} (1992) 165--168}.

\bibitem{Libanov:1995gh}
M.~V. Libanov, D.~T. Son and S.~V. Troitsky, \emph{{Exponentiation of
  multiparticle amplitudes in scalar theories. 2. Universality of the
  exponent}}, \href{https://doi.org/10.1103/PhysRevD.52.3679}{\emph{Phys. Rev.}
  {\bfseries D52} (1995) 3679--3687},
  [\href{https://arxiv.org/abs/hep-ph/9503412}{{\ttfamily hep-ph/9503412}}].

\bibitem{Witten:1998qj}
E.~Witten, \emph{{Anti-de Sitter space and holography}},
  \href{https://doi.org/10.4310/ATMP.1998.v2.n2.a2}{\emph{Adv. Theor. Math.
  Phys.} {\bfseries 2} (1998) 253--291},
  [\href{https://arxiv.org/abs/hep-th/9802150}{{\ttfamily hep-th/9802150}}].

\bibitem{Denner:2005nn}
A.~Denner and S.~Dittmaier, \emph{{Reduction schemes for one-loop tensor
  integrals}},
  \href{https://doi.org/10.1016/j.nuclphysb.2005.11.007}{\emph{Nucl. Phys.}
  {\bfseries B734} (2006) 62--115},
  [\href{https://arxiv.org/abs/hep-ph/0509141}{{\ttfamily hep-ph/0509141}}].

\bibitem{Passarino:1978jh}
G.~Passarino and M.~J.~G. Veltman, \emph{{One Loop Corrections for e+ e-
  Annihilation Into mu+ mu- in the Weinberg Model}},
  \href{https://doi.org/10.1016/0550-3213(79)90234-7}{\emph{Nucl. Phys.}
  {\bfseries B160} (1979) 151--207}.

\bibitem{Hahn:2000kx}
T.~Hahn, \emph{{Generating Feynman diagrams and amplitudes with FeynArts 3}},
  \href{https://doi.org/10.1016/S0010-4655(01)00290-9}{\emph{Comput. Phys.
  Commun.} {\bfseries 140} (2001) 418--431},
  [\href{https://arxiv.org/abs/hep-ph/0012260}{{\ttfamily hep-ph/0012260}}].

\bibitem{Hahn:1998yk}
T.~Hahn and M.~Perez-Victoria, \emph{{Automatized one loop calculations in
  four-dimensions and D-dimensions}},
  \href{https://doi.org/10.1016/S0010-4655(98)00173-8}{\emph{Comput. Phys.
  Commun.} {\bfseries 118} (1999) 153--165},
  [\href{https://arxiv.org/abs/hep-ph/9807565}{{\ttfamily hep-ph/9807565}}].

\bibitem{Contino:2016spe}
R.~Contino et~al., \emph{{Physics at a 100 TeV pp collider: Higgs and EW
  symmetry breaking studies}},
  \href{https://doi.org/10.23731/CYRM-2017-003.255}{\emph{CERN Yellow Report}
  (2017) 255--440}, [\href{https://arxiv.org/abs/1606.09408}{{\ttfamily
  1606.09408}}].

\bibitem{CMS:2013xfa}
{\scshape CMS} collaboration, \emph{{Projected Performance of an Upgraded CMS
  Detector at the LHC and HL-LHC: Contribution to the Snowmass Process}},  in
  \emph{{Proceedings, 2013 Community Summer Study on the Future of U.S.
  Particle Physics: Snowmass on the Mississippi (CSS2013): Minneapolis, MN,
  USA, July 29-August 6, 2013}}, 2013,
  \href{https://arxiv.org/abs/1307.7135}{{\ttfamily 1307.7135}},
  \href{http://inspirehep.net/record/1244669/files/arXiv:1307.7135.pdf}{http://inspirehep.net/record/1244669/files/arXiv:1307.7135.pdf}.

\bibitem{deBlas:2016ojx}
J.~de~Blas, M.~Ciuchini, E.~Franco, S.~Mishima, M.~Pierini, L.~Reina et~al.,
  \emph{{Electroweak precision observables and Higgs-boson signal strengths in
  the Standard Model and beyond: present and future}},
  \href{https://doi.org/10.1007/JHEP12(2016)135}{\emph{JHEP} {\bfseries 12}
  (2016) 135}, [\href{https://arxiv.org/abs/1608.01509}{{\ttfamily
  1608.01509}}].

\bibitem{No:2016ezr}
J.~M. No and M.~Spannowsky, \emph{{A Boost to $h \to Z \gamma$: from LHC to
  Future $e^+ e^-$ Colliders}},
  \href{https://doi.org/10.1103/PhysRevD.95.075027}{\emph{Phys. Rev.}
  {\bfseries D95} (2017) 075027},
  [\href{https://arxiv.org/abs/1612.06626}{{\ttfamily 1612.06626}}].

\bibitem{Amhis:2016xyh}
Y.~Amhis et~al., \emph{{Averages of $b$-hadron, $c$-hadron, and $\tau$-lepton
  properties as of summer 2016}},
  \href{https://arxiv.org/abs/1612.07233}{{\ttfamily 1612.07233}}.

\bibitem{Buras:1993xp}
A.~J. Buras, M.~Misiak, M.~Munz and S.~Pokorski, \emph{{Theoretical
  uncertainties and phenomenological aspects of B $\to$ X(s) gamma decay}},
  \href{https://doi.org/10.1016/0550-3213(94)90299-2}{\emph{Nucl. Phys.}
  {\bfseries B424} (1994) 374--398},
  [\href{https://arxiv.org/abs/hep-ph/9311345}{{\ttfamily hep-ph/9311345}}].

\bibitem{Ali:1990vp}
A.~Ali and C.~Greub, \emph{{A Profile of the final states in B $\to$ X(s) gamma
  and an estimate of the branching ratio BR (B $\to$ K* gamma)}},
  \href{https://doi.org/10.1016/0370-2693(91)90156-K}{\emph{Phys. Lett.}
  {\bfseries B259} (1991) 182--190}.

\bibitem{Chen:2001fja}
{\scshape CLEO} collaboration, S.~Chen et~al., \emph{{Branching fraction and
  photon energy spectrum for $b \to s \gamma$}},
  \href{https://doi.org/10.1103/PhysRevLett.87.251807}{\emph{Phys. Rev. Lett.}
  {\bfseries 87} (2001) 251807},
  [\href{https://arxiv.org/abs/hep-ex/0108032}{{\ttfamily hep-ex/0108032}}].

\bibitem{Limosani:2009qg}
{\scshape Belle} collaboration, A.~Limosani et~al., \emph{{Measurement of
  Inclusive Radiative B-meson Decays with a Photon Energy Threshold of
  1.7-GeV}}, \href{https://doi.org/10.1103/PhysRevLett.103.241801}{\emph{Phys.
  Rev. Lett.} {\bfseries 103} (2009) 241801},
  [\href{https://arxiv.org/abs/0907.1384}{{\ttfamily 0907.1384}}].

\bibitem{Lees:2012ym}
{\scshape BaBar} collaboration, J.~P. Lees et~al., \emph{{Precision Measurement
  of the $B \to X_s \gamma$ Photon Energy Spectrum, Branching Fraction, and
  Direct CP Asymmetry $A_{CP}(B \to X_{s+d}\gamma)$}},
  \href{https://doi.org/10.1103/PhysRevLett.109.191801}{\emph{Phys. Rev. Lett.}
  {\bfseries 109} (2012) 191801},
  [\href{https://arxiv.org/abs/1207.2690}{{\ttfamily 1207.2690}}].

\bibitem{Bennett:2006fi}
{\scshape Muon g-2} collaboration, G.~W. Bennett et~al., \emph{{Final Report of
  the Muon E821 Anomalous Magnetic Moment Measurement at BNL}},
  \href{https://doi.org/10.1103/PhysRevD.73.072003}{\emph{Phys. Rev.}
  {\bfseries D73} (2006) 072003},
  [\href{https://arxiv.org/abs/hep-ex/0602035}{{\ttfamily hep-ex/0602035}}].

\bibitem{Jegerlehner:2009ry}
F.~Jegerlehner and A.~Nyffeler, \emph{{The Muon g-2}},
  \href{https://doi.org/10.1016/j.physrep.2009.04.003}{\emph{Phys. Rept.}
  {\bfseries 477} (2009) 1--110},
  [\href{https://arxiv.org/abs/0902.3360}{{\ttfamily 0902.3360}}].

\bibitem{Hagiwara:2011af}
K.~Hagiwara, R.~Liao, A.~D. Martin, D.~Nomura and T.~Teubner, \emph{{(g-2)mu
  and alpha(MZ2) re-evaluated using new precise data}},
  \href{https://doi.org/10.1088/0954-3899/38/8/085003}{\emph{J. Phys.}
  {\bfseries G38} (2011) 085003},
  [\href{https://arxiv.org/abs/1105.3149}{{\ttfamily 1105.3149}}].

\bibitem{Davier:2010nc}
M.~Davier, A.~Hoecker, B.~Malaescu and Z.~Zhang, \emph{{Reevaluation of the
  Hadronic Contributions to the Muon g-2 and to alpha(MZ)}},
  \href{https://doi.org/10.1140/epjc/s10052-012-1874-8,
  10.1140/epjc/s10052-010-1515-z}{\emph{Eur. Phys. J.} {\bfseries C71} (2011)
  1515}, [\href{https://arxiv.org/abs/1010.4180}{{\ttfamily 1010.4180}}].

\bibitem{Hanneke:2010au}
D.~Hanneke, S.~F. Hoogerheide and G.~Gabrielse, \emph{{Cavity Control of a
  Single-Electron Quantum Cyclotron: Measuring the Electron Magnetic Moment}},
  \href{https://doi.org/10.1103/PhysRevA.83.052122}{\emph{Phys. Rev.}
  {\bfseries A83} (2011) 052122},
  [\href{https://arxiv.org/abs/1009.4831}{{\ttfamily 1009.4831}}].

\bibitem{Giudice:2012ms}
G.~F. Giudice, P.~Paradisi and M.~Passera, \emph{{Testing new physics with the
  electron g-2}}, \href{https://doi.org/10.1007/JHEP11(2012)113}{\emph{JHEP}
  {\bfseries 11} (2012) 113},
  [\href{https://arxiv.org/abs/1208.6583}{{\ttfamily 1208.6583}}].

\bibitem{Degrande:2016oan}
C.~Degrande, V.~V. Khoze and O.~Mattelaer, \emph{{Multi-Higgs production in
  gluon fusion at 100 TeV}},
  \href{https://doi.org/10.1103/PhysRevD.94.085031}{\emph{Phys. Rev.}
  {\bfseries D94} (2016) 085031},
  [\href{https://arxiv.org/abs/1605.06372}{{\ttfamily 1605.06372}}].

\bibitem{Itzykson:1980rh}
C.~Itzykson and J.~B. Zuber, \emph{{Quantum Field Theory}}.
\newblock International Series In Pure and Applied Physics. McGraw-Hill, New
  York, 1980.

\end{thebibliography}\endgroup

\end{document}